\documentclass{aa}
\usepackage[varg]{txfonts}
\usepackage{color}
\usepackage{natbib}
\usepackage{graphicx}	
\usepackage{amsmath}	
\usepackage{amssymb}	
\usepackage{multirow}   
\usepackage[draft]{hyperref}

\usepackage{ulem}     

\bibpunct{(}{)}{;}{a}{}{,} 

\newcommand{\mrm}[1]{\ensuremath\mathrm{#1}}
\newcommand{\GX}{GX~339$-$4}
\newcommand{\MJD}[1]{MJD~#1}

\begin{document}

\title{Colors and patterns of black hole X-ray binary GX 339$-$4} 


\author{Ilia A. Kosenkov\inst{1, 2} \and Alexandra Veledina\inst{1,3,4} \and Valery F. Suleimanov\inst{4,5,6} \and Juri Poutanen\inst{1,3,4}} 

\institute{
Department of Physics and Astronomy, FI-20014 University of Turku, Finland
\and Department of Astrophysics, St. Petersburg State University, Universitetskiy pr. 28, Peterhof, 198504 St. Petersburg, Russia
\and Nordita, KTH Royal Institute of Technology and Stockholm University, Roslagstullsbacken 23, SE-10691 Stockholm, Sweden
\and Space Research Institute of the Russian Academy of Sciences, Profsoyuznaya Str. 84/32, 117997 Moscow,  Russia
\and Institut f\"{u}r Astronomie und Astrophysik, Kepler Center for Astro and Particle Physics, Universit\"{a}t T\"{u}bingen, Sand 1, 72076 T\"{u}bingen, Germany
\and Kazan (Volga region) Federal University, Kremlevskaya str. 18, Kazan 420008, Russia
} 
\date{}

\abstract{
    Black hole X-ray binaries show signs of non-thermal emission in the optical/near-infrared range.
    We analyze the optical/near-infrared SMARTS data on \GX~over the 2002--2011 period. 
    Using the soft state data, we estimate the interstellar extinction towards the source and characteristic color temperatures of the accretion disk.
    We show that various spectral states of regular outbursts occupy similar regions on the color-magnitude diagrams, and that transitions between the states proceed along the same tracks despite substantial differences in the observed light curves morphology. 
    We determine the typical duration of the hard-to-soft and soft-to-hard state transitions and the hard state at the decaying stage of the outburst to be  one, two and four weeks, respectively.
    We find that the failed outbursts cannot be easily distinguished from the regular ones at their early stages, but if the source reaches 16 mag in $V$-band, it will transit to the soft state.
    By subtracting the contribution of the accretion disk, we obtain the spectra of the non-thermal component, which have constant, nearly flat shape during the transitions between the hard and soft states. 
    In contrast to the slowly evolving non-thermal component seen at optical and near-infrared wavelengths, the mid-infrared spectrum is strongly variable on short timescales and sometimes shows a prominent excess with a cutoff below $10^{14}$~Hz.
    We show that the radio to optical spectrum can be modeled using three components corresponding to the jet, hot flow and irradiated accretion disk.
   }
    
\keywords{accretion, accretion disks -- black hole physics -- methods: data analysis -- stars: black holes --  stars: individual: GX 339$-$4 -- X-rays: binaries }
\maketitle 
    
\section{Introduction}
    
\object{GX~339--4}~is a well-studied black hole (BH) low-mass X-ray binary (LMXB) that was discovered as a bright and variable X-ray source using MIT \textit{OSO--7} experiment \citep{Markert1973}. 
The source is known to undergo recurrent outbursts every 2--3 years and it became a standard target for multiwavelength campaigns \citep{HBMB05,Belloni2005,Belloni2006,Tomsick2008,Shidatsu2011,CB11,Motta2011,Rahoui2012,Buxton2012,Corbel2013}.
During outbursts the source undergoes a transition between different states that can be distinguished using a variety of criteria, including the X-ray hardness ratio, quasi-periodic oscillations and other timing properties \citep{Homan2005a, McClintock2006, Remillard2006, Belloni2010}.

The origin of these states is still not very well understood, but is generally associated with the changes in the accretion flow geometry \citep[see, e.g.,][]{Esin1997,Poutanen1997,Zdziarski2004,Done2007}. 
In the soft X-ray state, the spectrum consists of a thermal component associated with the standard cold accretion disk \citep{SS73} and an additional power-law-like tail from the non-thermal corona \citep{PC98,Gier99,ZGP01}. 
In the hard state, the emission from the cold disk is greatly reduced due to its truncation at a radius significantly larger than the radius of the innermost stable orbit; the emission instead is dominated by thermal Comptonization of some seed photons (cold disk and internal synchrotron) in an inner hot, geometrically thick  accretion flow. 
The transitions between the spectral states happen due to a change in the truncation radius, which results in a variation of the relative contributions of the cold disk and the hot flow and also lead to corresponding changes in the timing properties  \citep{Done2007,Gilfanov10,Poutanen2014a,DeMarco2015, Stiele2017, Poutanen2018,  Mahmoud2019}. 

For many years it was obvious that the X-ray emitting region of accreting BHs is rather compact because of the observed fast variability. 
On the other hand, the optical/near-infrared  (ONIR)  emission was thought to originate mostly from the outer disk irradiated by the central X-ray source.
The first evidence that the situation is not so simple came already in the beginning of 1980s, when the fast optical variability and quasi-periodic oscillations at 20\,s were detected from \GX~ by  \citet{Motch82}, which were interpreted as a signature of emission from the hot accretion flow or corona \citep{Fabian82}. 
Soon after,  \citet{Motch83} carried out simultaneous optical/X-ray observations which demonstrated a  complicated structure of the cross-correlation function (CCF) with a precognition dip.
Recently, similar CCFs were found in three BHs: XTE~J1118+480 \citep{Kanbach01,HHC03}, Swift~J1753.5$-$0127 \citep{DGS08,DGS09,DSG11,HBM09} and in \GX~\citep{GMD08,Gandhi2010}. 
The shape of the CCF can be explained if the  optical emission contains  two components: synchrotron emission from the hot flow and the reprocessed radiation that are anti-correlated  and  correlated with the X-rays, respectively \citep{VPV11}.  
However, the shape of the CCF seems to be wavelength-dependent. 
For instance, the infrared/X-ray CCF of \GX~showed a single peak with no precognition dip \citep{CMO10}, which was successfully explained by a jet model, where the emission is powered by internal shocks \citep{Malzac2018}. 
Thus it is clear that the ONIR emission is not completely dominated by the accretion disk, but has at least one additional (non-thermal) component, either from the hot flow, or the jet, or both. 

\begin{figure}
\centering
    \includegraphics[keepaspectratio, width = 0.9\linewidth]{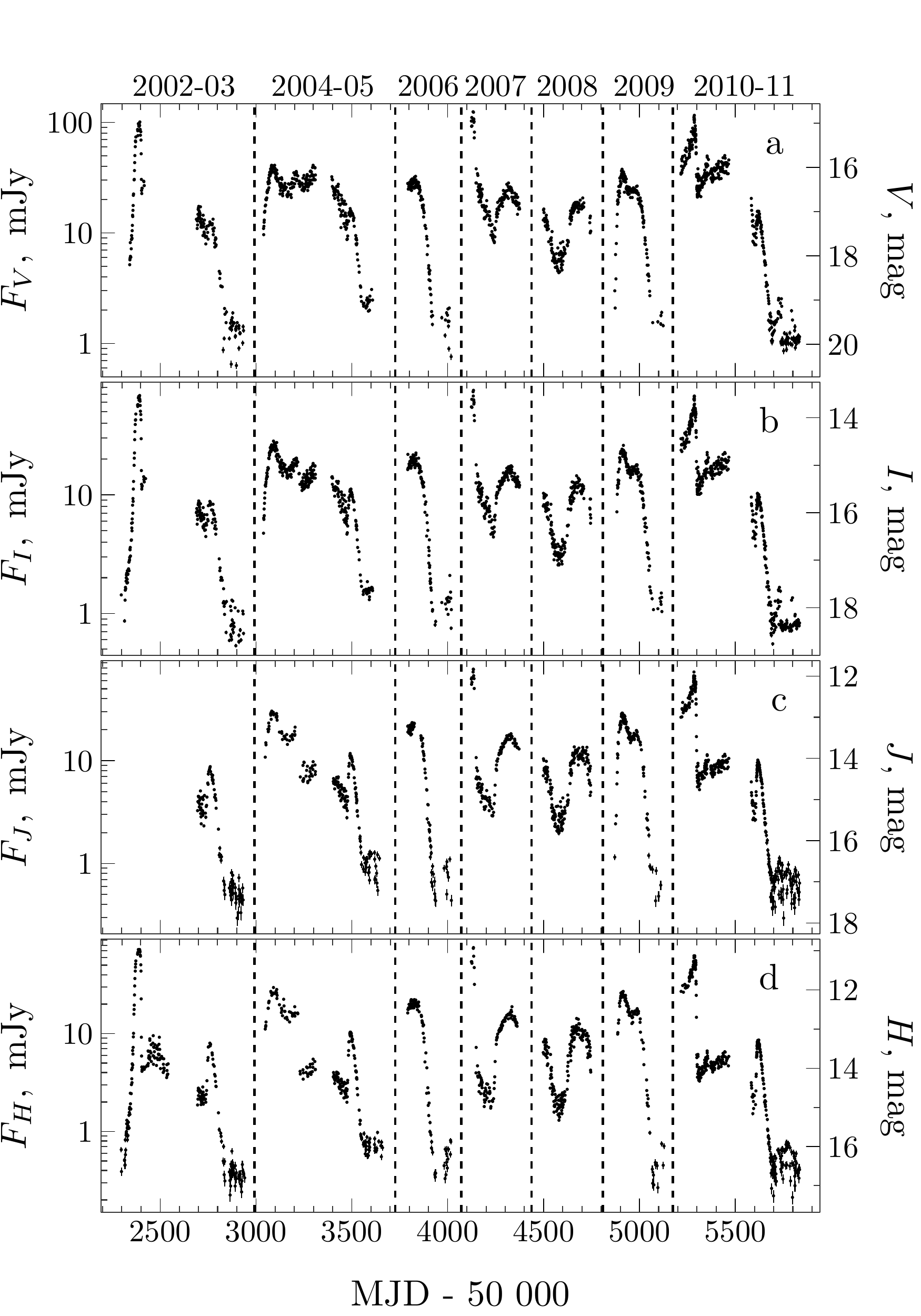}
    \caption{
        Observed ONIR magnitudes and corrected for the interstellar extinction fluxes of \GX~in four ONIR filters ($V$, $I$, $J$ and $H$).
        Vertical dashed lines separate data into 7 intervals corresponding to each of the regular/failed outbursts.}
    \label{fig:LC}
\end{figure}

Signatures of this component are  seen in the long-term variability as bright ONIR flares, which appear in the hard state \citep{Hynes2000,JBOM01, BB04, KDT13}.
Spectral evolution of the flare in XTE~J1550--564 is consistent with the hot accretion flow scenario  \citep{Veledina2013, Poutanen2014a}. 
In Swift~J1753.5--0127 the jet emission is known to be weak in radio and its contribution to the ONIR band is likely small \citep{DGS09}, while the ONIR-X-ray broad-band spectra can well be explained by the hot flow model \citep{Kajava2016}.
On the other hand, the soft ONIR spectra of 4U~1543--47 \citep{Kalemci2005} and MAXI~J1836--194 \citep{Russell13} argue in favor of the jet scenario. 

The origin of this non-thermal component can give us a clue to understanding the nature of the accretion-ejection engine operating in the BH vicinity. 
This aim can be reached by studying the evolution of its spectral shape throughout the outburst and in particular during the transitions between the states.
This requires measurements of the interstellar extinction and subtraction of the (irradiated) accretion disk contribution.
From this perspective, \GX~is an interesting source because it has served as a target for numerous multiwavelength campaigns. 
The aim of this work is to study the behavior of the source spectrum on the color-magnitude diagram (CMD) and the evolution of the non-thermal component using spectral decomposition. 

The paper is structured as follows.   
In Sect.~\ref{sec:Data}, we describe the data used for the analysis. 
Sect.~\ref{sec:results} contains details of the analysis, determination of the interstellar extinction,  comparison of different outbursts, and determination of the spectrum of the non-thermal component. 
In Sect.~\ref{sec:discussion}  we discuss the observed ONIR properties and their connection to the X-rays.  
We also discuss the origin of the non-thermal component in terms of the hot flow and jet models. 
We conclude in Sect.~\ref{sec:conclusions}. 

\begin{table*}
    \centering
    \caption{Descriptions of the outburst phases.}
    \label{tbl:spec_states}
    \begin{tabular}{lcp{10cm}}
    \hline
    \hline
    Outburst phase & Abbreviation &  Description \\
    \hline
    Rising phase & RPh & The initial phase of the outburst, during which both X-ray and ONIR fluxes increase and the source becomes redder \\
    Rising hard state & RHS & The period in the first ONIR flare, during the hard X-ray state, when the ONIR colors stabilize\\
    Hard to soft transition & HtS & The short, about 5--10~d, transition to the soft state, over which the ONIR colors become bluer and X-ray spectra become softer \\
    Soft state & SS & Period during the soft X-ray state when the ONIR spectra are dominated by the accretion disk emission \\
    Soft to hard transition & StH & The short, about 10--20 d, period over which the ONIR colors become redder and the source departs from the soft X-ray state \\
    Decaying hard state & DHS & The period within the second flare of about 20--30~d, when the ONIR colors are stable and red \\
    Decaying phase & DPh & The period when ONIR fluxes decay towards the quiescence \\
    Quiescence & Q & The period of low activity between the outbursts \\
    \hline
    \end{tabular}
\end{table*}

\begin{table*}
    \centering
    \caption{
        Start and end dates of the outbursts and start dates of each identified outburst phase (in MJD). }
    \label{tbl:intrvl}
    \begin{tabular}{cllllllllll}
        \hline
        \hline
          Years    & Start        &  RPh & RHS  &  HtS  &  SS  &  StH  &  DHS  &  DPh  & Q  & End \\
                  \hline
          \multicolumn{11}{c}{Regular outbursts}    \\
          2002--2003  & 52298     & 52298 & 52374 & 52400 & 52405 & 52738 & 52756 & 52786 & 52834  & 52942 \\
           2004--2005  & 53040      & & 53040 & 53221 & 53231 & 53473 & 53490 & 53513 & 53553 & 53661 \\
          2007        & 54124     & & 54124 & 54137 & 54146 & 54236 & 54252 &  &  & 54376 \\
         2010--2011  & 55217    & & 55217& 55295 & 55302 & 55602 & 55613 & 55645 & 55689  &55836  \\
        \hline
                 \multicolumn{11}{c}{Failed outbursts}    \\
     2006        & 53792      &&&&&& 53792 & 53867 & 53935 & 54020 \\          
     2008        & 54498     && 54498 &   & &   & 54634 &&  & 54747 \\
     2009        & 54872      & 54872 & 54885 &&&  & 54953 & 55000 & 55083 & 55129\\   
       \hline
    \end{tabular}
        \tablefoot{Phase definitions are described in  Table~\ref{tbl:spec_states} and Section~\ref{sec:separation}.}  
\end{table*}

\section{Data}
\label{sec:Data}

\begin{figure}
\centering
    \includegraphics[keepaspectratio, width = 0.9\linewidth]{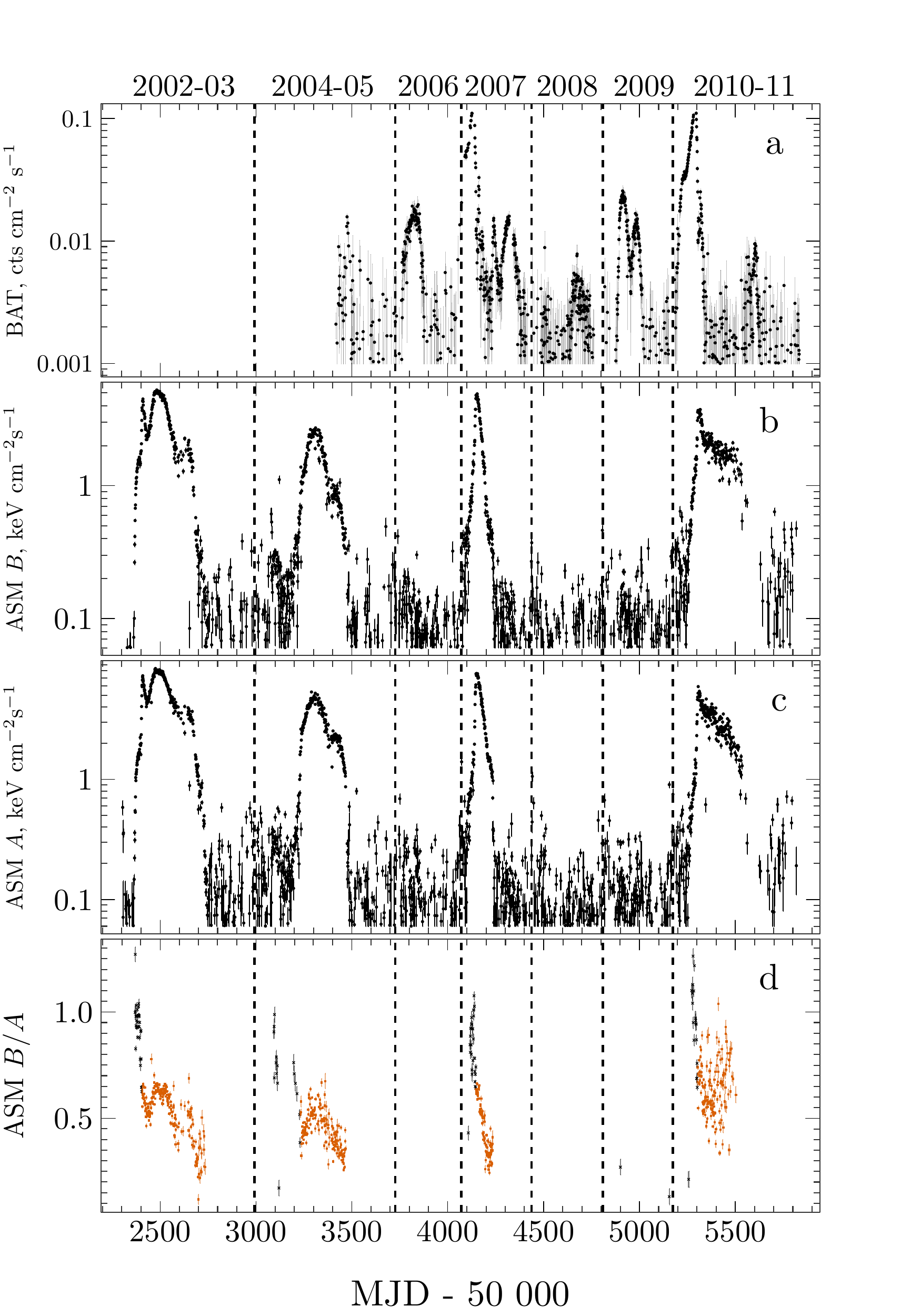}
    \caption{X-ray  light curves of \GX: (a) \textit{Swift}/BAT count rate in the 15--150 keV band, (b) ASM\,B (3--5 keV) flux, (c) ASM\,A (1.5--3 keV) light curve, (d) ASM\,B/A hardness ratio. 
    Orange points correspond to the soft states selected using ONIR data from four outbursts. Errors are 1$\sigma$.}
    \label{fig:XR}
\end{figure}

Monitoring of \GX~in the ONIR has been conducted using the ANDICAM camera \citep{DePoy2003} on the Small and Moderate Aperture Research Telescope System \citep[SMARTS;][]{Subasavage2010}.\footnote{\url{http://www.astro.yale.edu/smarts/xrb/home.php}}
We use the publicly available {SMARTS} data in $V$, $I$, $J$ and $H$ filters taken in 2002--2011 \citep{Buxton2012}.\footnote{The data mentioned with the original publication \citep{Buxton2012} cover 2002--2010 period and  the fluxes in $I$ filter are computed using an incorrect extinction $A_I$. }
We split the ONIR light curves into 7 intervals, each covering one of the regular or failed outbursts (see Sect.~\ref{sec:separation}).
The light curves, corrected for interstellar extinction (see Sect.~\ref{sec:extinction}), are shown in Fig.~\ref{fig:LC}.

The X-ray data are taken from the \textit{Rossi X-ray Timing Explorer} All-Sky Monitor (ASM)\footnote{\url{http://xte.mit.edu/ASM_lc.html}} \citep{Bradt1993}, which provides observations of \GX~in three energy ranges (1.5--3, 3--5  and 5--12 keV). 
The count rates from different channels are converted to the observed fluxes using an empirical linear relation of \citet{Zdziarski2002}. 
We also make use of the \textit{Neil Gehrels Swift Observatory} Burst Alert Telescope (\textit{Swift}/BAT)\footnote{\url{https://swift.gsfc.nasa.gov/results/transients/weak/GX339-4/}} \citep{Gehrels2004, Barthelmy2005}, which operates in the 15--150~keV range. 
The BAT, ASM\,A and B light curves as well as the ASM\,B/A hardness ratio are shown in Fig.~\ref{fig:XR}.

The mid-infrared (mid-IR) data, obtained by the \textit{Wide-field Infrared Survey Explorer} satellite \citep[\textit{WISE},][]{Wright2010}, are adopted from \citet{Gandhi2011}. 
All reported 13  observations of \GX\ are taken during 24 hours on \MJD{55265.88--55266.88}, i.e. within half a day from the SMARTS observation on \MJD{55266.36} \citep{Buxton2012}. 
The mid-IR data are corrected for the interstellar extinction \citep[see section~2.2.3 in][]{Gandhi2011}.

We also use the radio data obtained by the Australian Telescope Compact Array \citep[ATCA,][]{Wilson2011} in 5.5 and 9~GHz bands.
Out of two closest to the SMARTS/\textit{WISE} quasi-simultaneous spectrum observations \citep[\MJD{55262.91~and~55269.80}, see table~1 in][]{Corbel2013} we select the second one, which has a positive slope in the radio and can be extrapolated to the mid-IR \textit{WISE} data.

\section{Light curve analysis and results}
\label{sec:results}
 
\subsection{Outburst phase separation}
\label{sec:separation}

We use ONIR light curves to split regular outbursts, when \GX~reaches the soft state,  into different phases: rising phase (RPh), 
hard state at the rising stage (RHS), transition from the hard to the soft state (HtS), soft state (SS), transition from the soft to the hard state (StH), hard state  at the decaying stage (DHS), decaying phase (DPh) and quiescence (Q), see Table~\ref{tbl:spec_states}.
Different states are first roughly identified using the X-ray data, as in \citet{Motta2009}.
We then fit an S- or Z-shaped curves to the ONIR data during the RPh, HtS and StH phases and determine the transition dates.
Further, we fit the broken power-law to the DPh and Q data and obtain the start dates of quiescence. 
The details of the described procedure are explained in Appendix~\ref{Appendix} and the start dates of each outburst phase (if this phase has been observed) are presented in Table~\ref{tbl:intrvl}. 
We also apply this fitting procedure to the failed outbursts, for which we identify ONIR phases with colors, similar to the phases of the regular outbursts. 
 
Our definition of phases can be different from those in other works. 
For instance, the ONIR StH transitions defined in \citet{KDT13} occur 2--6 days later than ours (compare our Table~\ref{tbl:intrvl} to their table~2). 
At the same time, their transition dates defined by the X-ray spectral and timing properties are up to 20 days earlier.
However, our start dates of the RHS and HtS of the 2007 outburst agree well with the transition dates obtained in \citet{Motta2009} using the X-ray spectral data.

\begin{table}
    \centering
    \caption{ONIR filter effective wavelengths $\lambda$, zero-point fluxes $F_0$  \citep{Buxton2012} and derived extinction.  }
    \label{tbl:filters}
    \begin{tabular}{crrl}
        \hline
        \hline
        Filter & $\lambda$  & $F_0$   &  $A_\lambda$  \\
        & ($\AA$) &  (Jy)  &   (mag) \\
        \hline
        $V$    &    5450   &  3636   & 3.58    \\
        $I$    &    7980   &  2416   & 2.16 (+0.15)   \\
        $J$    &   12500   &  1670   & 1.00   \\
        $H$    &   16500   &   980   & 0.64   \\
        \hline
    \end{tabular}
    \tablefoot{Extinction coefficients are calculated using interstellar extinction laws from \citet{Cardelli89} and \citet{ODonnell94}. }
\end{table}

\subsection{Determination of extinction}
\label{sec:extinction}

\begin{figure}
    \includegraphics[keepaspectratio, width=0.9\linewidth]{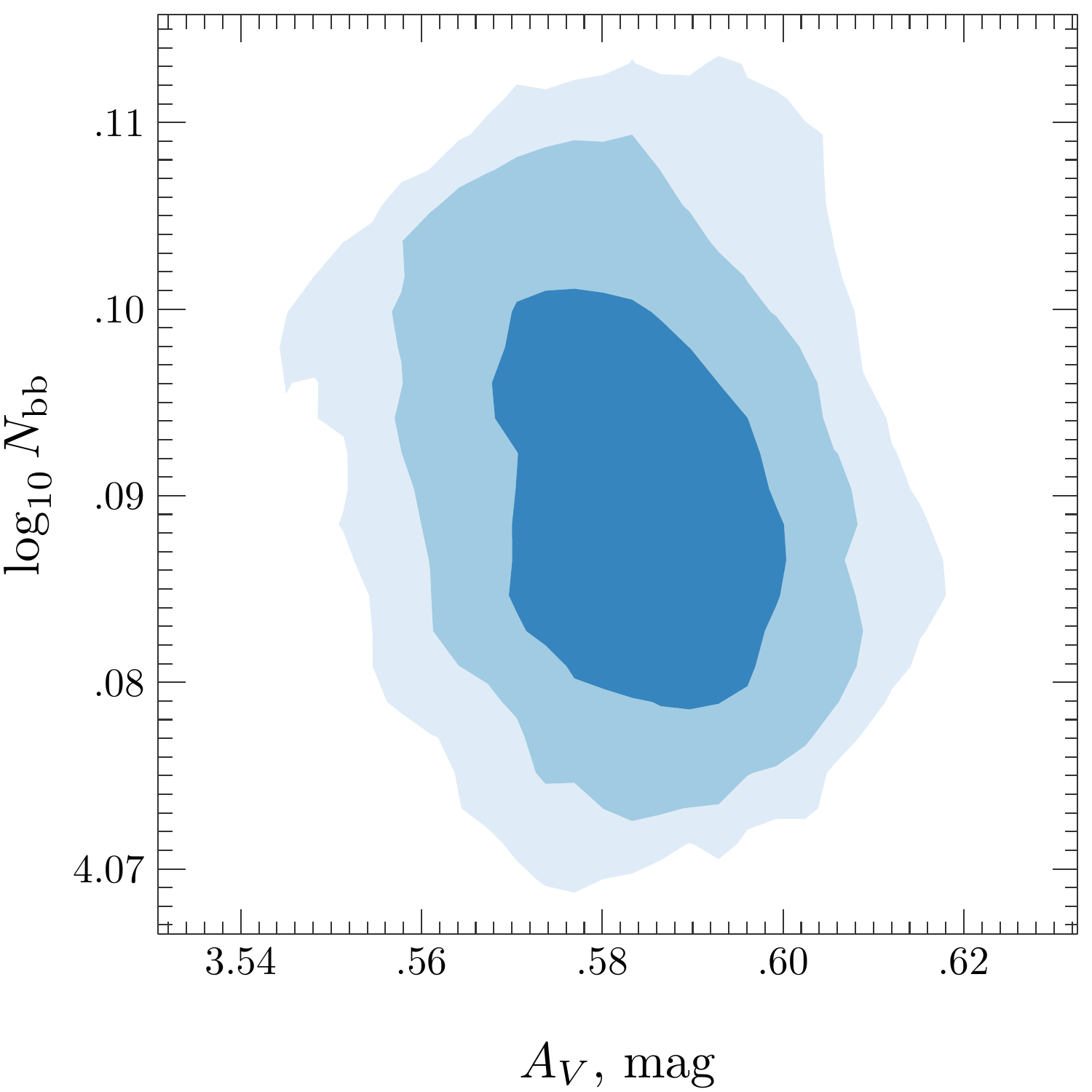}
    \caption{Joint posterior distribution of the model parameters. Contours correspond to 0.68, 0.95 and 0.99 probabilities. }
    \label{fig:av_vs_n}
\end{figure}

We use the SS data from four regular outbursts to determine the extinction $A_V$ towards the source.
We select 236 nights, during which \GX\ was observed in all four ONIR filters. 
We convert magnitudes to fluxes using the zero-points given in \citet[][see also Table 2]{Buxton2012}.  
Assuming that the SS spectra can be described by the blackbody model, we fit the data  with the diluted Planck function modified by interstellar extinction: 
\begin{equation}
    \label{eq:BB_FIT}
    F_{k,j} = 10^{-A_k / 2.5}\  N_{\rm bb}\  B_k(T_j),
\end{equation}
where $k$ corresponds to one of the ONIR filter ($V$, $I$, $J$, $H$), index $j$ corresponds to the date, $F_{k,j}$ is the model flux (in mJy), $B_k$ is the value of the Planck function at the effective frequency of filter $k$, $T_j$ is the estimated blackbody temperature, $A_k$ is the extinction (in magnitudes) in $k$-filter obtained using the model of \citet{Cardelli89} with correction by \citet{ODonnell94}.
The dimensionless normalization $N_{\rm bb}$ is assumed to be the same for all outbursts. 
It can be expressed through the effective radius of the irradiated disk  $R_\mrm{irr}$, the distance to the source $D$ and the disk inclination $i$ as: 
\begin{equation}
    \label{eq:NORM}
N_{\rm bb} =  5.09 \times 10^{4}\ \pi \left(\frac{R_\mrm{irr}/R_\odot}{D\,(\rm kpc)}\right)^2  \cos i\ .
\end{equation}

We noticed that the fluxes in the $I$ filter are systematically offset from the blackbody fits for any $A_V$. 
A possible reason can be deviation of the irradiated disk spectrum from a simple blackbody. 
Therefore, for the fitting purposes, we introduce a correction $\Delta A_I$  to the extinction in the $I$ filter as an additional parameter.  
The best-fit parameters estimated using Bayesian inference are $A_V = 3.58 \pm 0.02$\,mag, $\Delta A_I  = 0.15 \pm 0.02$\,mag and $\log_{10} N_{\rm bb} = 4.09 \pm 0.01$.
Fig.~\ref{fig:av_vs_n} shows the joint posterior distribution of the fitted $A_V$ and $N_{\rm bb}$. 

Extinction can also be estimated using the color excess $E_{B-V}$. 
The value $E_{B-V} = 0.94 \pm 0.19$\,mag was derived from the X-ray absorption by \citet{HBMB05}, while \citet{Zdziarski1998} give $E_{B-V} = 1.2 \pm 0.1$\,mag as a weighted mean obtained from different methods. 
Using interstellar absorption lines \citet{Buxton2003} got  $E_{B-V} = 1.1 \pm 0.2$\,mag.
These extinction estimates correspond to $A_V$ between 2.9  and 3.7\,mag \citep[assuming $R_V = 3.1$,][]{Cardelli89}. 
Thus, our estimate of $A_V$ is well within the range of previously used values. 

Our assumption that the blackbody normalization remains the same for all outbursts can be violated.
For example, the effective disk radius in the LMXB XTE~J1859+226 decreased with decreasing disk temperature \citep{Hynes2002}. 
It is unclear whether this happens due to the changes in the geometrical size of the disk or disk warping, or because of the propagation of the cooling wave \citep{King1998}. 
\GX~shows complex behavior in its soft states (see Section~\ref{sec:regular_outbursts} for the discussion), so it is possible that the normalization is indeed variable, which can systematically offset the inferred value of $A_V$.
To address this issue we tried to fit $T_j$ and $N_{{\rm bb},j}$ individually for each night, with $A_V$ and $\Delta A_I$ remaining global parameters.
This problem is numerically unstable, because in the Rayleigh-Jeans regime  (i.e. at high temperatures) Eq.~(\ref{eq:BB_FIT}) transforms into
\begin{equation}
    F_{k,j} = 10 ^{-A_k/2.5} \frac{2\nu_k^2 k_{\rm B}}{c^2} N_{{\rm bb},j} T_j.
\end{equation}
As a result, contributions of $N_{{\rm bb},j}$ and $T_j$ cannot be easily decoupled and observations at higher energies (e.g. in UV) are required to properly fit temperatures and normalizations.

We also investigated the influence of the variable effective radius $R_\mrm{irr}$ on the fitted value of $A_V$ by introducing different normalizations for each outburst. 
We find no evidence that the variation in the normalization   significantly affects the inferred value of $A_V$.

\begin{table}[]
    \centering
        \caption{Parameters of the system.}
        \label{tbl:orbital}
        \begin{tabular}{llr@{$~\pm~$}l@{~}l}
            \hline
            \hline
            Mass ratio                            & $q$             & 0.18    & 0.05\tablefootmark{a}   &            \\
            Orbital period                        & $P$             & 1.76    & 0.01\tablefootmark{a}                &  d         \\
            Black hole mass                       & $M_1$           & 5.9     & 3.6\tablefootmark{a}                 &  $M_\odot$ \\
            Inclination                           & $i$             & 58      & 21\tablefootmark{a}                &  deg       \\
            Distance                              & $D$             & 8       & 2\tablefootmark{b}                     &  kpc       \\
            Binary separation                     & $a$             & 11.6    & 5.1                                                    & $R_\odot$  \\
            Roche lobe size                       & $R_\mrm{L,1}$   & 6.1     & 3.2                                                    & $R_\odot$  \\
            Maximum disk size                     & $R_\mrm{tidal}$ & 5.9     & 2.6                                                    & $R_\odot$  \\
            Effective radius of irradiated disk   & $R_\mrm{irr}$   & 3.0     & 1.1                                                    & $R_\odot$ \\
             \hline
        \end{tabular}
        \tablefoot{
\tablefoottext{a}{From \citet{Heida2017}.}
\tablefoottext{b}{From \citet{Zdziarski2004}.}
}
\end{table}

\begin{figure}
\centering
    \includegraphics[keepaspectratio, width = 0.9\linewidth]{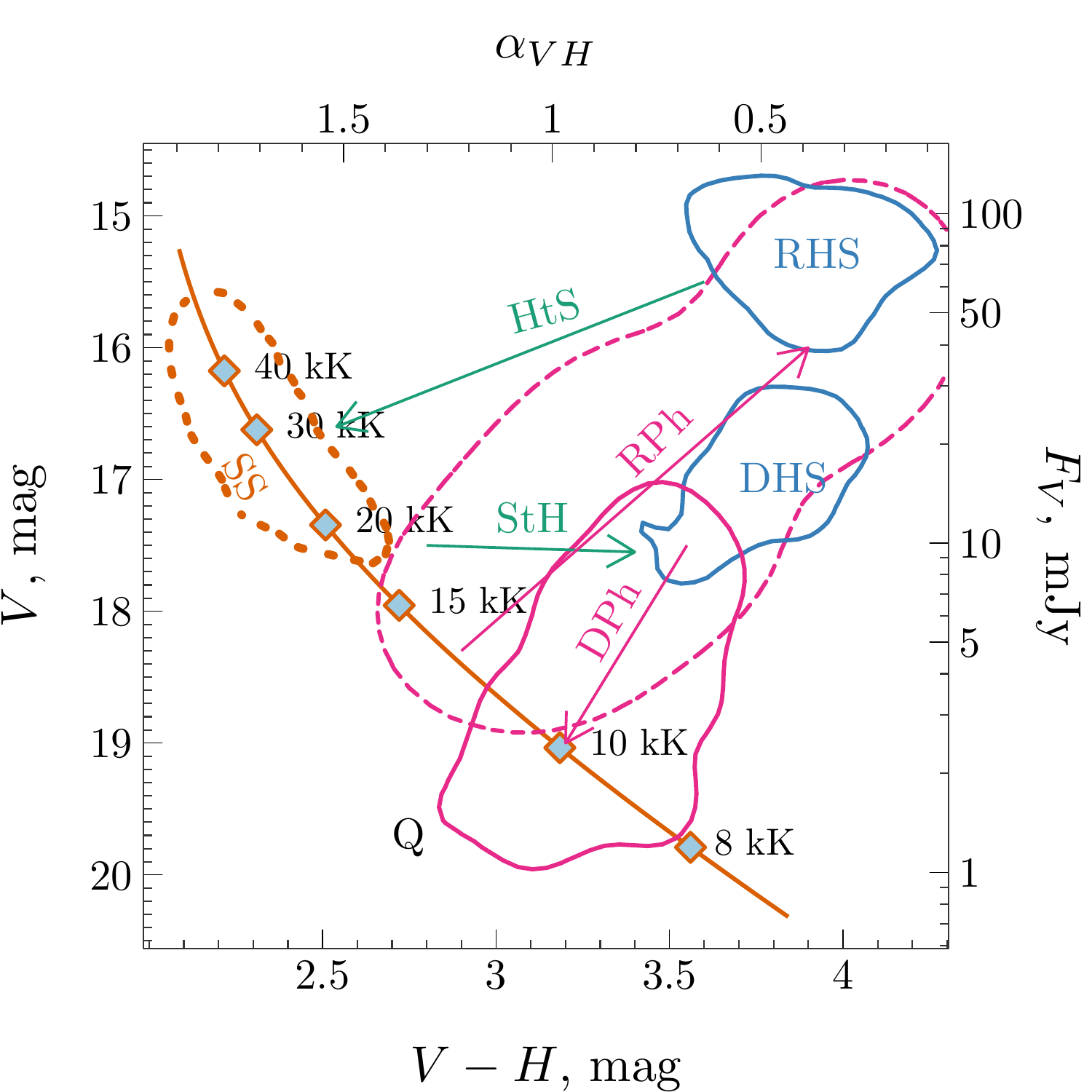}
    \caption{
    Evolution of magnitudes and colors of \GX~throughout the outburst. 
    $V$ and $V-H$ are observed magnitudes and colors, $F_V$ is corrected for the interstellar extinction flux and $\alpha_{VH}$ is the intrinsic spectral slope, computed assuming  $A_V = 3.58$\,mag (see Eq. \ref{eq:extin_coeff_short}).
    Dashed pink contour corresponds to the RPh, top-right solid blue contour -- to the RHS, top solid green arrow -- to the HtS transition, dotted orange contour -- to the SS, bottom solid green arrow -- to  the reverse StH transition, bottom solid blue contour -- to the DHS, solid pink contour -- to the DPh. 
    Solid orange line gives the model blackbody curve with $\log_{10} N_{\rm bb} = 4.09$, see Eq.\,(\ref{eq:BB_FIT}) and filled blue diamonds along the curve correspond to the temperatures marked on the right.
    The quiescent state is marked with letter Q in the lower part of the diagram. 
    }
    \label{fig:TEMPL}
\end{figure}

Using the obtained value of the blackbody normalization $N_{\rm bb}$ we now can estimate the characteristic size of the accretion disk. 
From the distance and the inclination of the binary system given in Table~\ref{tbl:orbital}, we obtain $R_\mrm{irr} = \left(3.0 \pm 1.1 \right)~{R}_\odot$.
We then estimate the binary separation $a / R_\odot = 4.17 (M_1/M_\odot)^{1/3} (1+q)^{1/3} P_\mrm{day}^{2/3}$ \citep[see e.g.][]{AccretionPower}, size of the primary Roche lobe $R_\mrm{L,1} / a = 0.49/\left[0.6  + q^{2/3}\log (1 + q^{-1/3})\right]$ \citep{Eggleton1983} and the maximum disk size due to the tidal instability $R_\mrm{tidal} / a = 0.60 / (1 + q)$ \citep{Pazcynski1977,CataclysmicVariableStars}.  
This allows us to compare the parameters of \GX~to its ``twin'' system XTE~J1550--564 \citep{MunozDarias2008}.

\begin{figure*}
\centering
       \includegraphics[keepaspectratio, width = 0.45\linewidth]{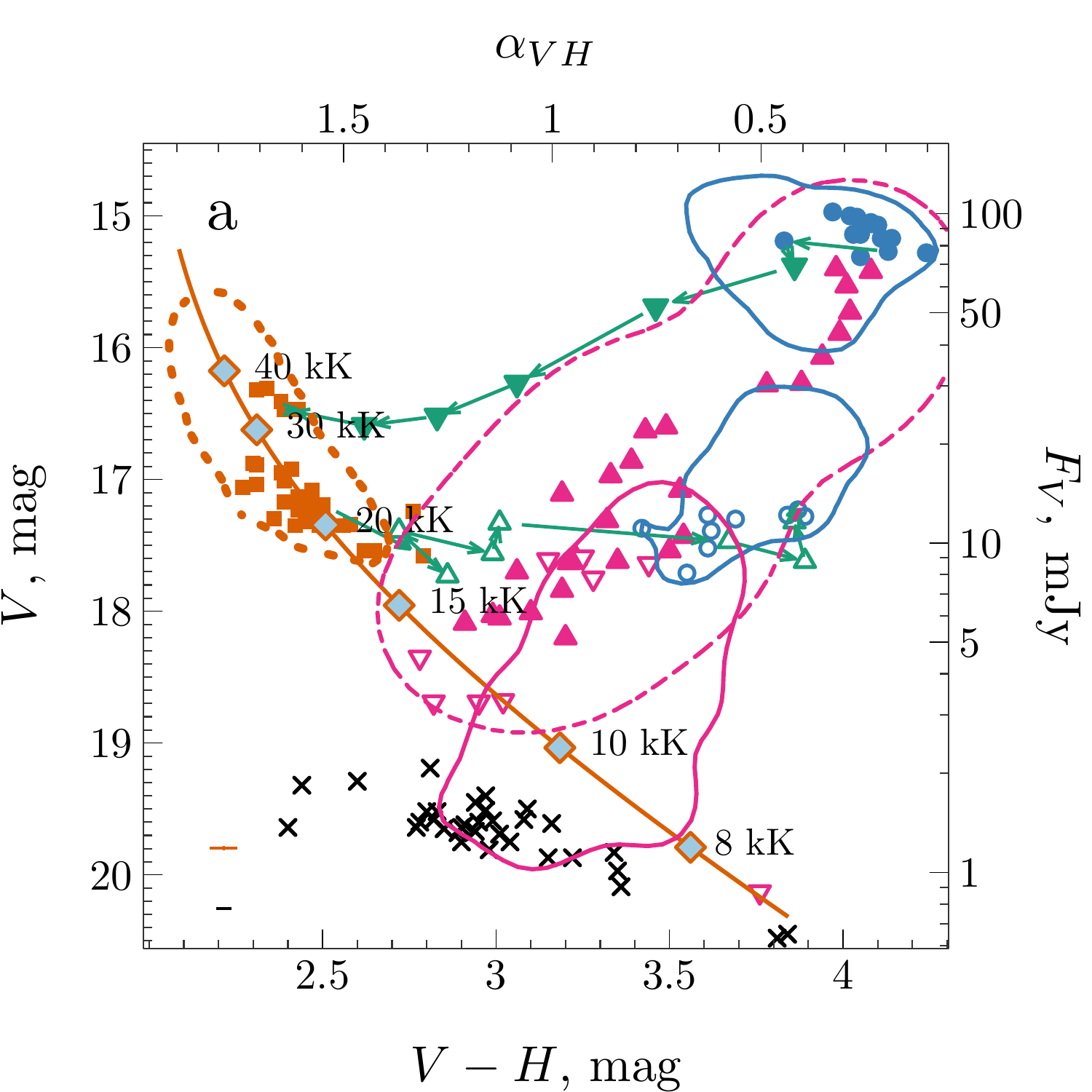}
        \includegraphics[keepaspectratio, width = 0.45\linewidth]{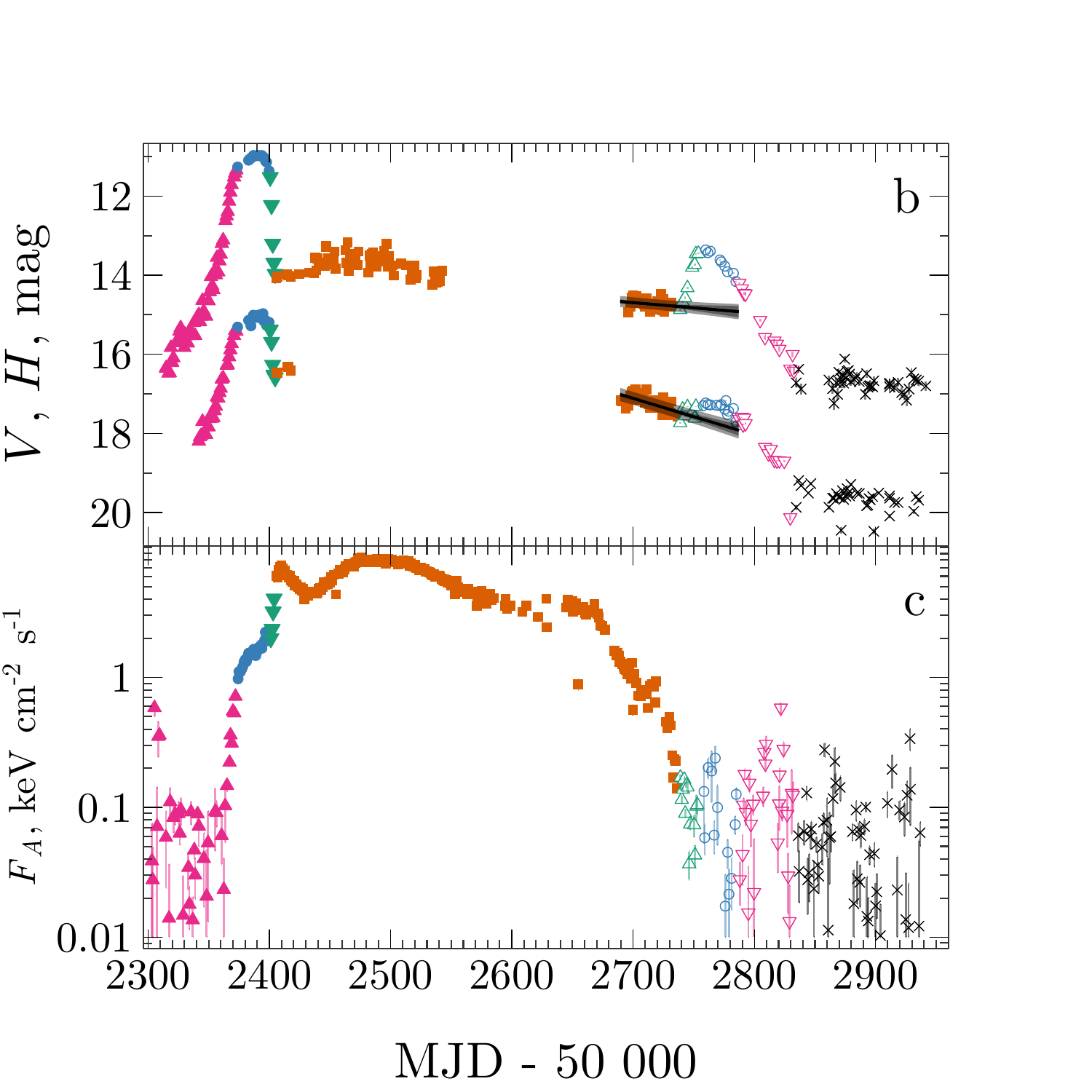}
   \caption{
        (a) $V$ versus $V-H$ color-magnitude diagram of the 2002--2003 outburst. 
        Filled upward-facing pink triangles correspond to the RPh, filled blue circles - to the RHS, filled downward-facing green triangles - to the HtS transition, filled orange squares -- to the SS, open upward-facing green triangles -- to the StH transition, open blue circles - to the DHS, open downward-facing pink triangles -- to the DPh, black crosses -- to the Q. 
        Contours are the same as in Fig.~\ref{fig:TEMPL}.
         The typical 1$\sigma$ errors are shown in the lower left corner: the top orange cross corresponds to the bright states and is comparable to the plot symbol size, the bottom black cross corresponds to the quiescent state data, where observational errors are substantial.
         (b) ONIR $H$ (top) and $V$ (bottom) light curves of the same outburst.
         Symbols are the same as in panel (a). 
         Errors are 1$\sigma$ and error bars are comparable to the symbols size. 
         Solid black line represents the model, fitted to the part of the SS data. 
         The grey area around the line denotes 1$\sigma$ errors and also includes the contribution from the intrinsic scatter. 
         (c) ASM\,A X-ray light curve. 
         }
    \label{fig:CMD_LC_01}
\end{figure*}

Analysis of the ONIR light curves of the 2000 outburst of \object{XTE~J1550--564} \citep{Poutanen2014} showed that the effective radius of the irradiated disk of XTE~J1550--564, $R_{\rm irr} \approx 4.1~{R}_\odot$, is 40\% smaller than the maximum disk size of 6.9~$R_\odot$. 
Although \GX~and XTE~J1550--564 binary systems have similar sizes \citep{Orocz2011, Hynes2003, Heida2017}, the major difference is the binary mass ratio $q$, which is of the order of 0.033 for XTE~J1550--564 and 0.18 for \GX~(see Table~\ref{tbl:orbital}).  
Larger mass ratio leads to a smaller primary Roche lobe size, limiting the maximum size of the accretion disk.

\subsection{Outburst template}

In this section we study the $V$--$(V-H)$ CMD of the regular outbursts to construct the outburst template. 
We use data from a given phase (RPh, RHS, SS, DHS and DPh) to compute the 90 per cent density contour (with the exception of the 2004--2005 outburst, whose RHS we omit because it is systematically fainter than that of the other regular outbursts). 
The obtained template is shown in Fig.~\ref{fig:TEMPL} and we further use it to study the similarities between regular outbursts and the peculiarities of the failed ones.

We compute the intrinsic spectral slopes $\alpha_{ij}$ (where the flux depends on frequency as $F_{\nu}\propto\nu^{\alpha}$), from the observed colors as
\begin{equation}
    \label{eq:extin_coeff}
    \alpha_{ij} = \frac{m_i - m_j - (A_i - A_j) - 2.5 \log\left(F_i^0 / F_j^0\right) }{2.5\log\left(\lambda_i/\lambda_j\right)}  ,
\end{equation}
where $i, j$ correspond to ONIR filters ($V$, $I$, $J$ or $H$), $m_i - m_j$ is the observed color, $A_i$ and $A_j$ are the interstellar extinctions, $\lambda_i$ and $\lambda_j$ are the effective wavelengths, and $F_i^0$ and $F_j^0$ are the zero-point fluxes in $i$ and $j$ filters, respectively (see Table~\ref{tbl:filters}).  
Equation (\ref{eq:extin_coeff}) for $V$ and $H$ filters can be reduced to 
\begin{equation}
        \label{eq:extin_coeff_short}
        \alpha_{VH} = -0.84\ (V-H) + 0.69\ A_V + 1.16.
\end{equation}

The observed RPh begins with the spectrum resembling blackbody with temperature of $T\approx 14$~kK. 
The spectrum then reddens as it enters the RHS, reaching $V-H \approx 4$\,mag.
During the HtS transition which takes 5--10~d, the spectrum becomes bluer.
This evolution may be attributed to the quenching of the red non-thermal component.
In the SS the source moves along the blackbody track (see Fig.~\ref{fig:TEMPL}, solid orange curve) towards upper-left part of the CMD, indicating a temperature increase from roughly 30 up to 50~kK. 
In 100--200~d, ONIR fluxes decrease and the temperature drops substantially. 
This marks the end of the SS and the beginning of the StH  transition, which occurs at $T\approx20$~kK. 
The StH transition takes 11--18~d and the source moves almost horizontally on the $V$--$(V-H)$ CMD, indicating an increase of flux in $H$ filter at almost no changes in $V$. 
\GX~spends 20--30~d in the DHS (although, in 2007 it lasted for more than 100~d) and after that it enters the DPh. 
During the next $\sim$50~d, the observed ONIR fluxes drop dramatically and the spectra become bluer.
The Q phase does not follow the blackbody track and shows a substantial (up to 1.5 mag) variability of $V-H$ color, unlike in other systems \citep[see][]{KDT13, Poutanen2014}.

We find that the colors and magnitudes of RHS are systematically different from those of DHS. 
The RPh is somewhat bluer than the DPh for the same $V$ magnitude (Fig.~\ref{fig:TEMPL}). 
The SS contour (solid orange line) is elongated along the blackbody track, but there is some spread of the data points around the model. 
This is likely caused by the variability due to superhumps \citep{Kosenkov2018}. 

The spread of colors in the Q phase is  peculiar. 
It cannot be explained by the measurement errors (see Fig.~\ref{fig:CMD_LC_01}a, bottom left corner), as it exceeds the typical error by a factor of 10. 
Interestingly, most of the Q phase data points lie below the blackbody line, i.e. the colors are typically bluer than the blackbody that has the same $V$-flux. 
We also note that the lowest observed blackbody temperature when the source enters the Q-phase is still above the hydrogen ionization limit.

\begin{figure*}
\centering
       \includegraphics[keepaspectratio, width = 0.45\linewidth]{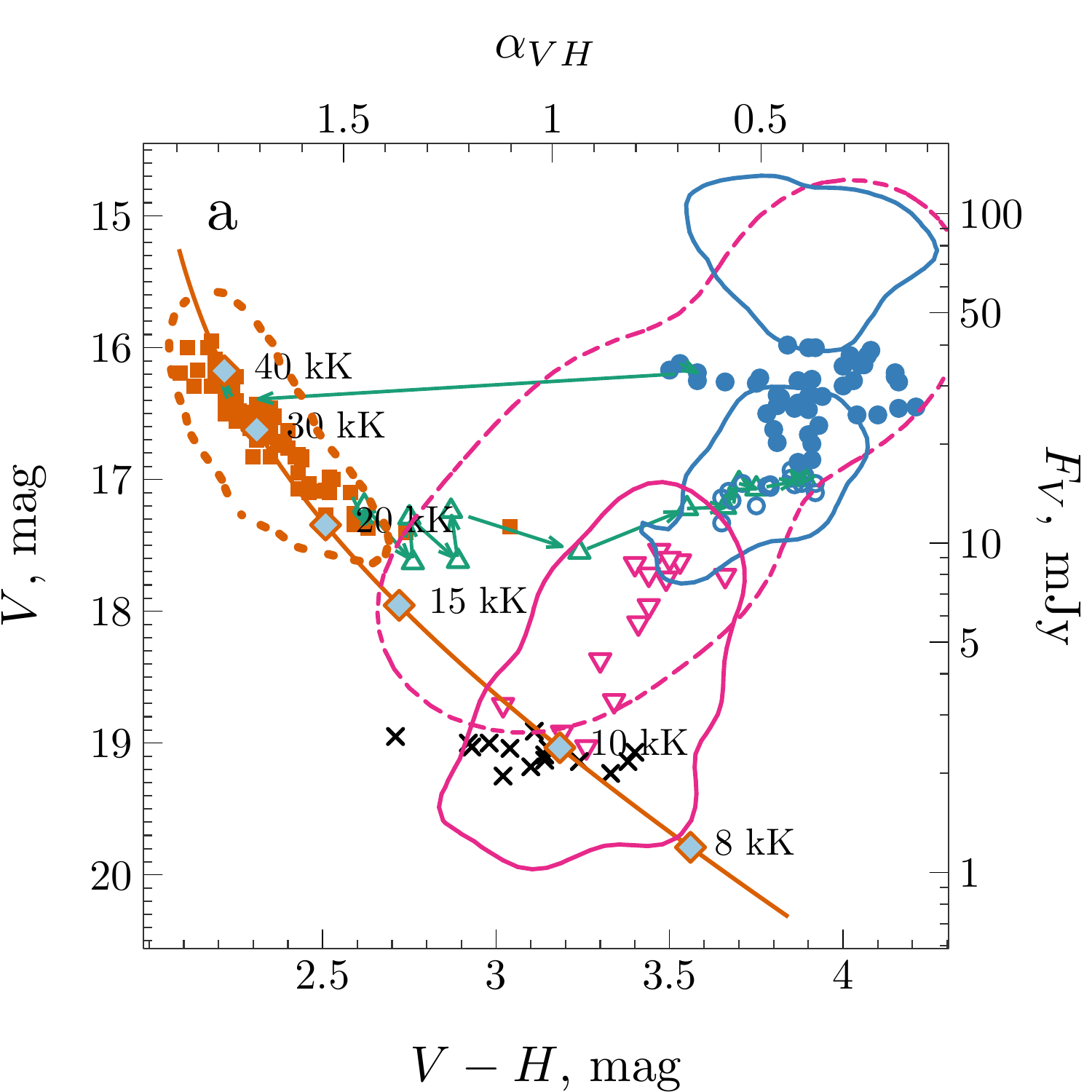}
      \includegraphics[keepaspectratio, width = 0.45\linewidth]{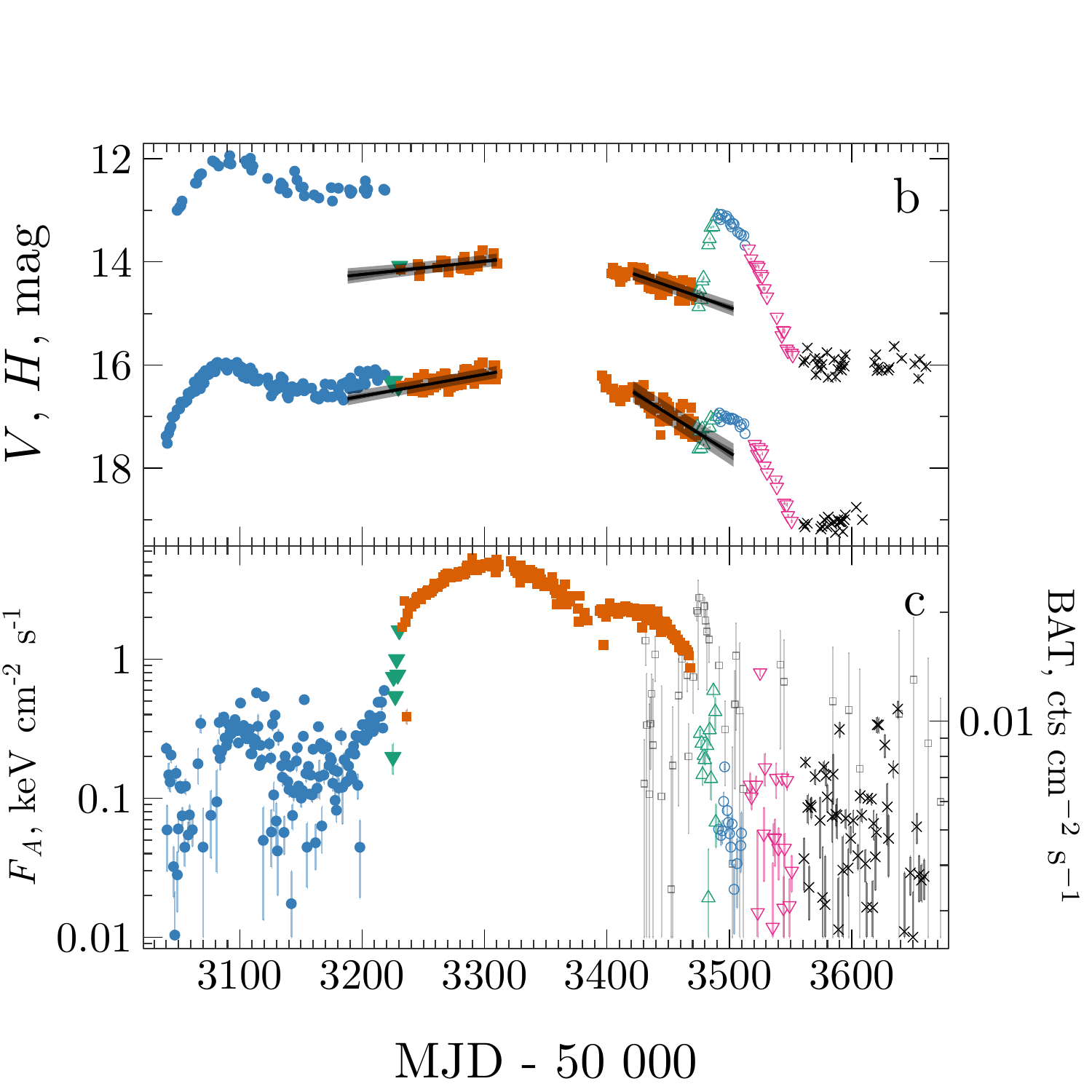}
   \caption{Same as Fig.~\ref{fig:CMD_LC_01}, but for the 2004--2005 outburst. 
    (c)  The BAT data are shown with open gray squares starting from the second half of the 2004--2005 outburst. Errors are 1$\sigma$.}
    \label{fig:CMD_LC_02}
\end{figure*}

\subsection{Regular outbursts}
\label{sec:regular_outbursts}

In this section we describe the properties of each outburst.
We stress the peculiarities and deviations from the common pattern discussed above. 
The 2002--2003 outburst is the only one with many observations of the initial RPh (filled pink upward-facing triangles in Fig.~\ref{fig:CMD_LC_01}). 
This outburst was also observed during the HtS transition (filled green downward-facing triangles). 
Thus, we can trace the evolution of \GX~before and after the RHS (filled blue circles) and compare it to the decay stages: StH transition (open green triangles),  DHS (open blue circles) and DPh (open pink triangles). 
Interestingly, after the HtS transition, the flux in $H$ filter showed no visible daily-timescale variability \citep[see Fig.~\ref{fig:CMD_LC_01}b; also][]{HBMB05}. 
In about 40~d (\MJD{52440}), when the flux started to decrease, the variability, which was previously attributed to superhumps \citep{Kosenkov2018}, became apparent. 
At the time of this change, the ASM fluxes and ASM\,B/A hardness ratio had a local minimum (see Fig.~\ref{fig:CMD_LC_01}b,c).

\begin{figure*}
\centering
       \includegraphics[keepaspectratio, width = 0.45\linewidth]{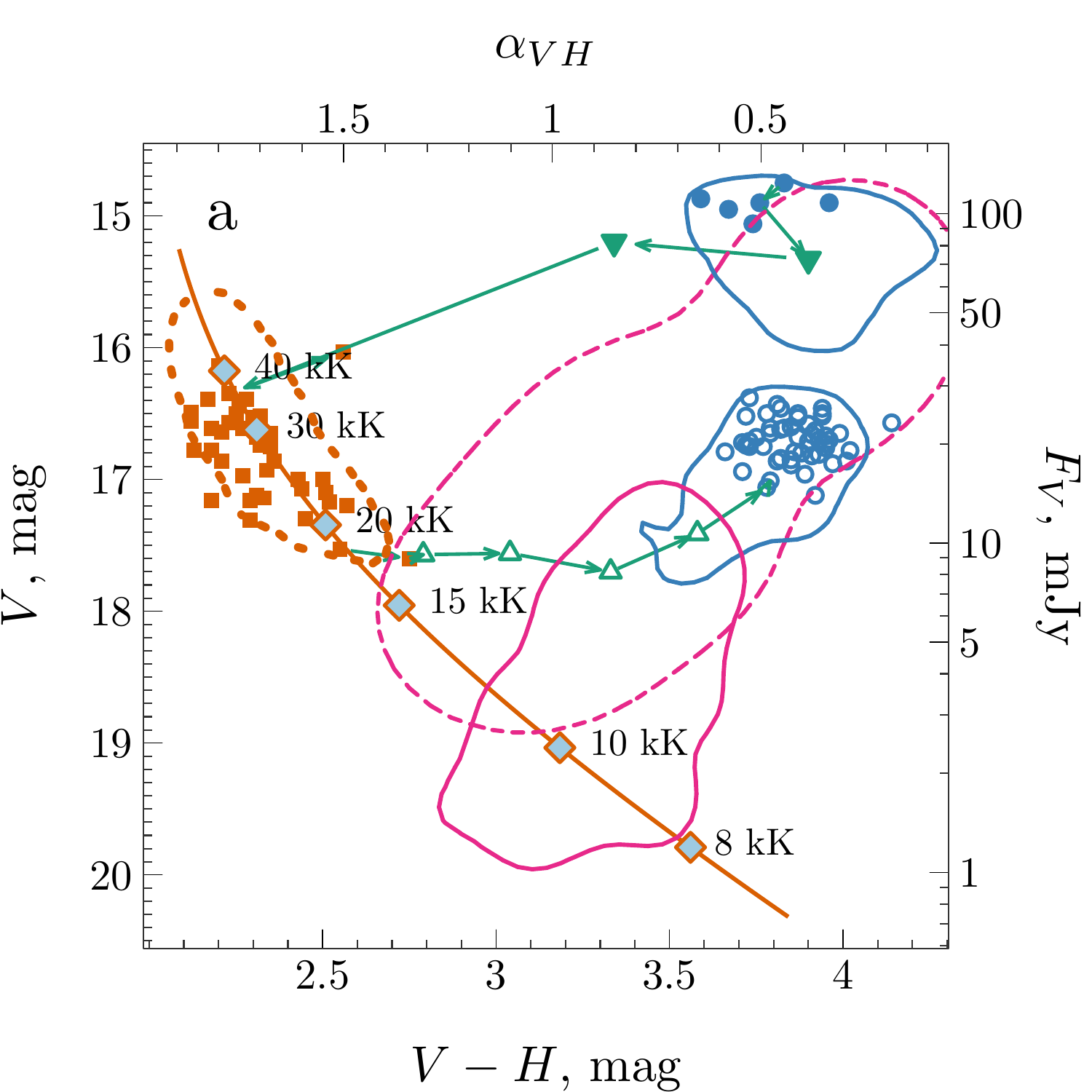}
      \includegraphics[keepaspectratio, width = 0.45\linewidth]{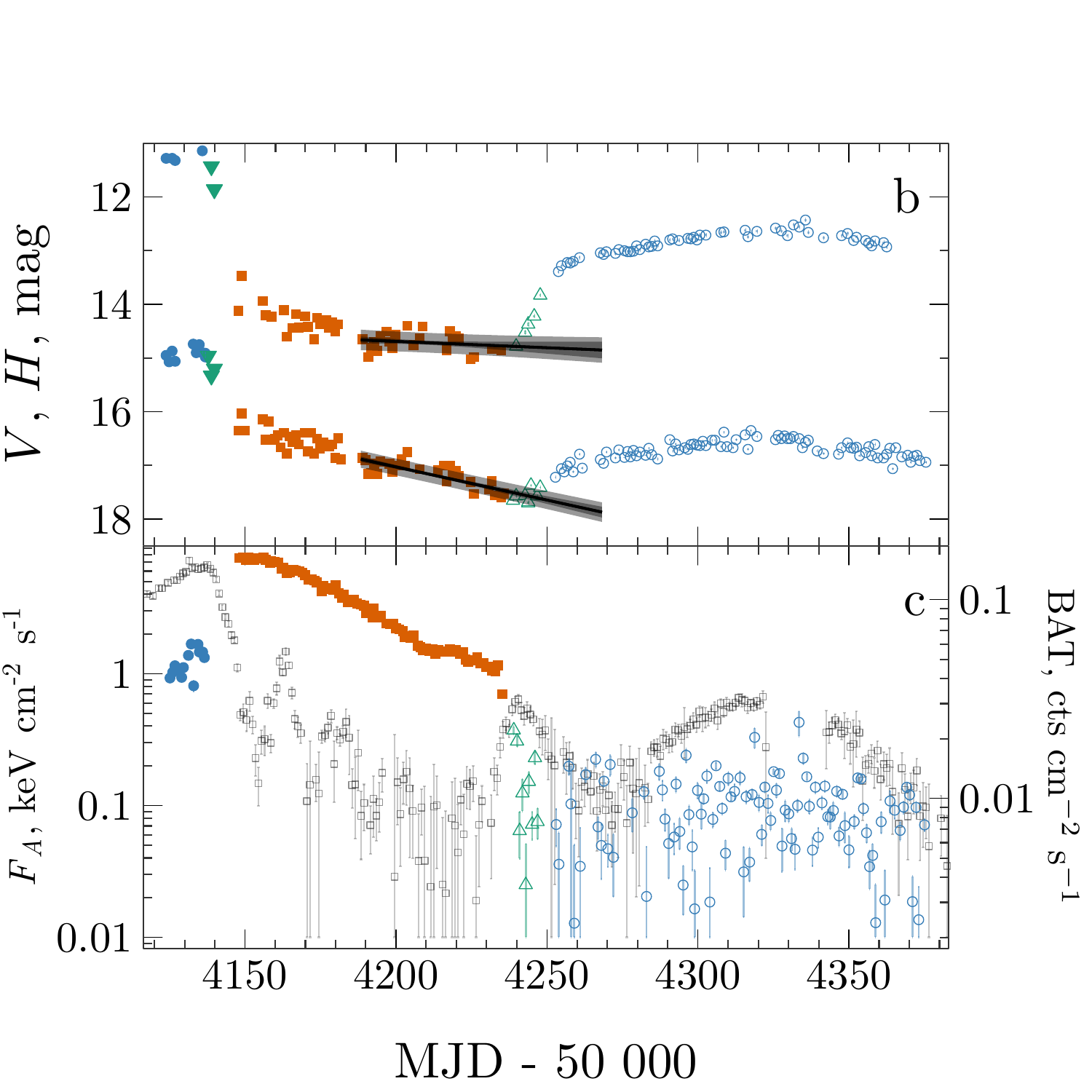}
   \caption{Same as Fig.~\ref{fig:CMD_LC_02}, but for the 2007 outburst.}
    \label{fig:CMD_LC_03}
\end{figure*}

\begin{figure*}
\centering
       \includegraphics[keepaspectratio, width = 0.45\linewidth]{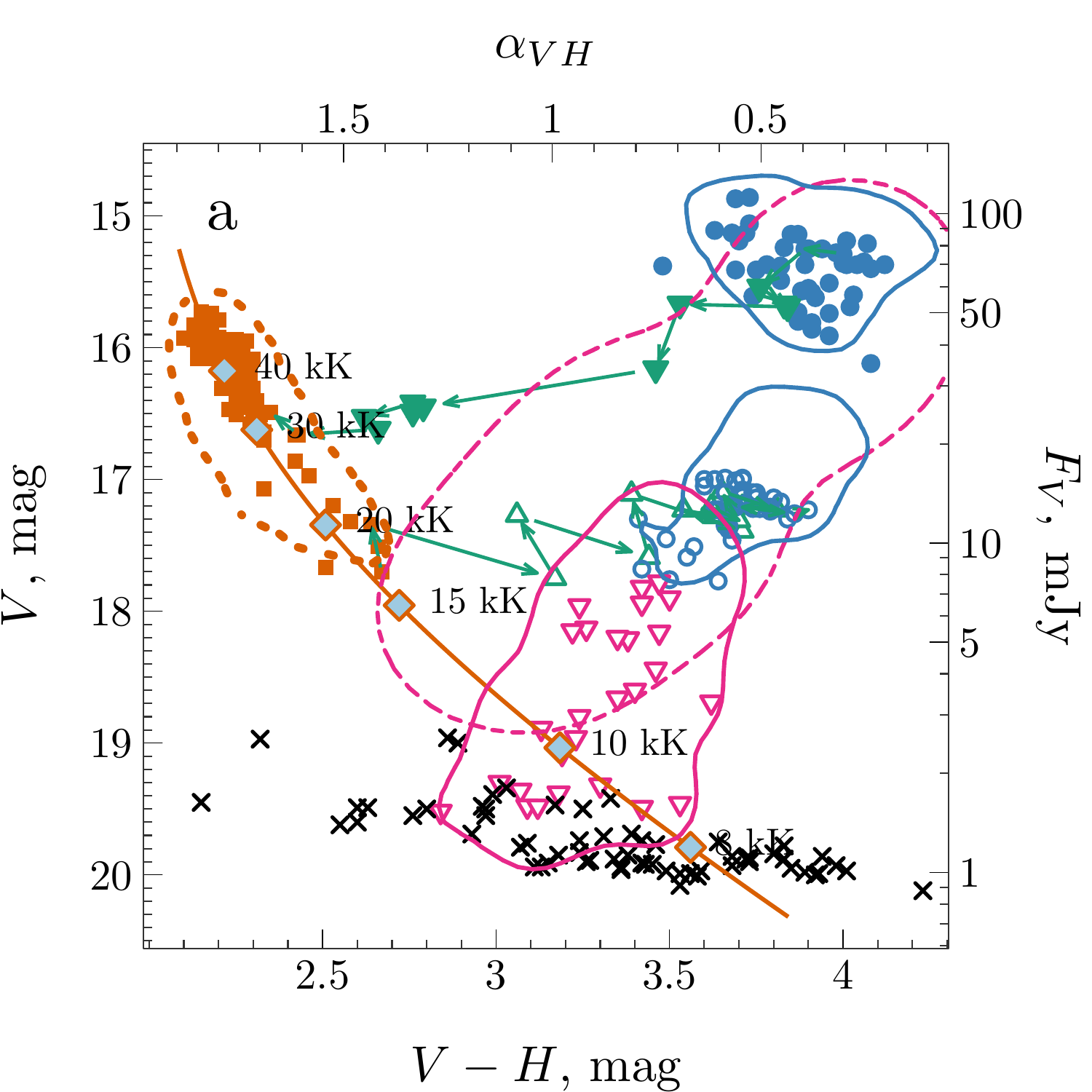}
      \includegraphics[keepaspectratio, width = 0.45\linewidth]{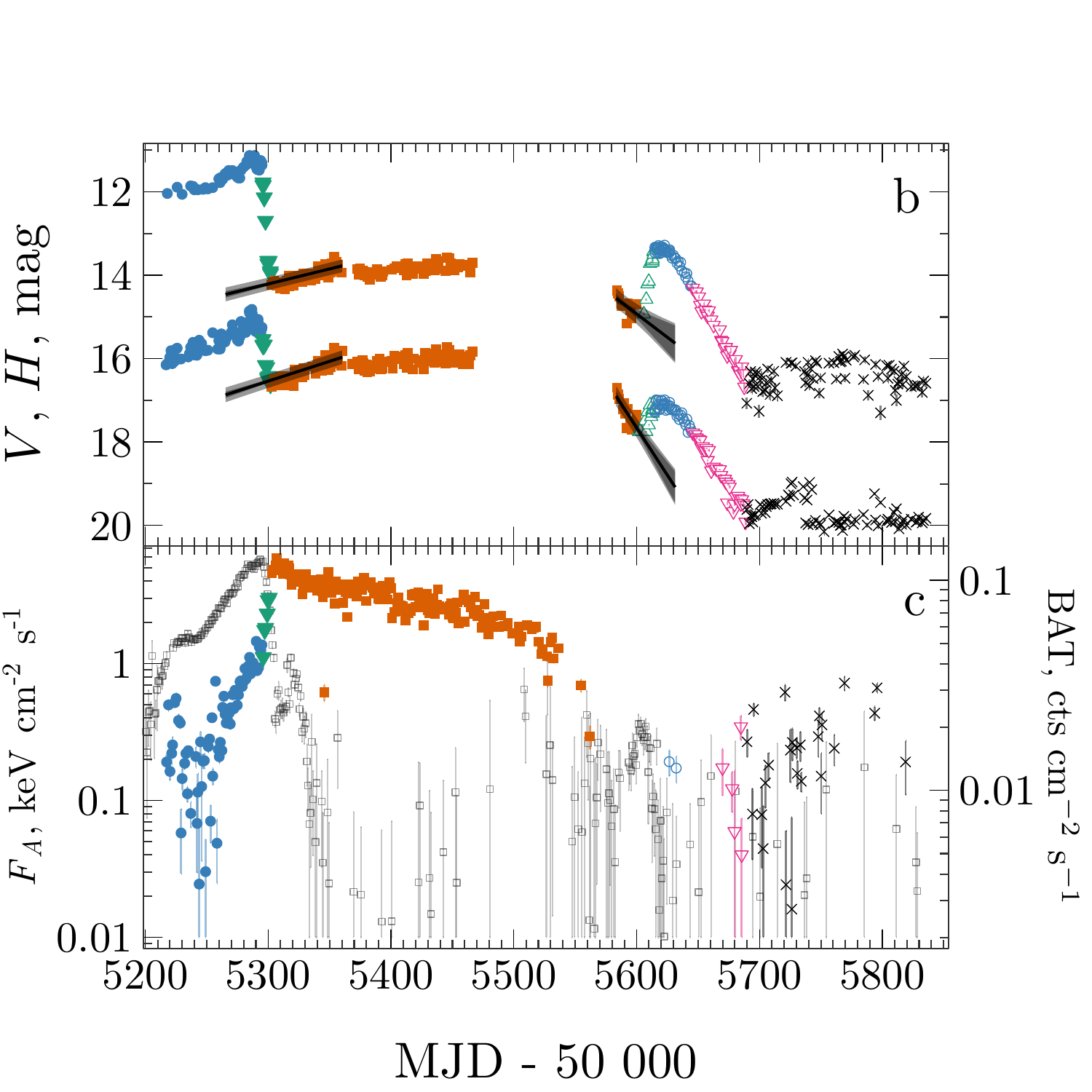}
   \caption{Same as Fig.~\ref{fig:CMD_LC_02}, but for the 2010--2011 outburst.}
    \label{fig:CMD_LC_04}
\end{figure*}

The 2004--2005 outburst was fainter, both in the X-rays and in ONIR, as compared to other regular outbursts (see Figs.~\ref{fig:LC} and \ref{fig:XR}). 
The same conclusion can be reached from Fig.~\ref{fig:CMD_LC_02}a, where the RHS (filled blue circles) are located between two template (RHS and DHS) contours (blue solid lines). 
The duration of the RHS, $\approx$180~d, is longer than that in the other outbursts (20--75~d).
The $V$-filter flux increased after the transition to SS, and then decreased together with the decreasing  ASM fluxes (Fig.~\ref{fig:CMD_LC_02}b,c). 
The DHS points are located well within the template contour, and the corresponding fluxes are only marginally lower than those of the RHS, in contrast to other regular outbursts.

\begin{figure*}
\centering
       \includegraphics[keepaspectratio, width = 0.45\linewidth]{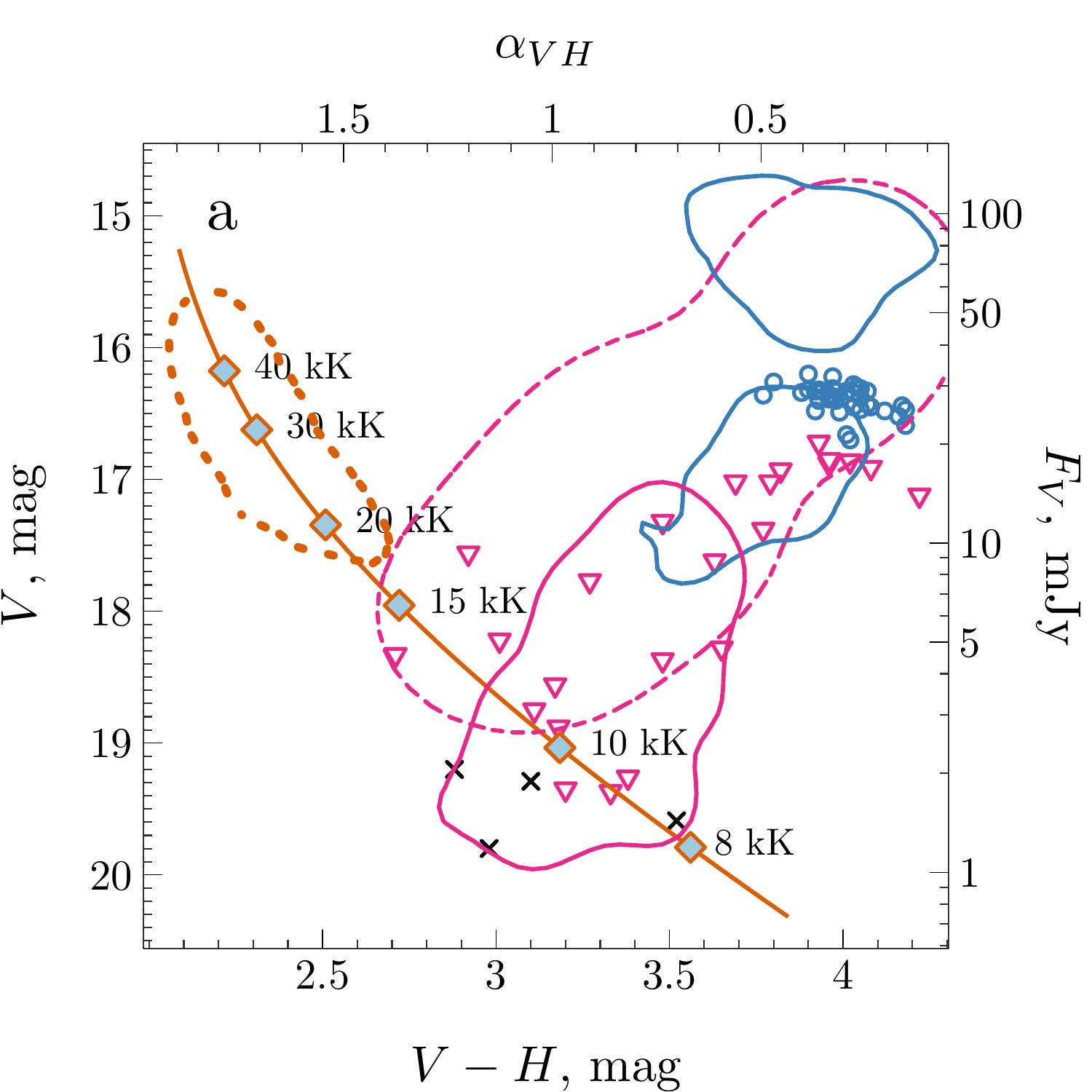}
      \includegraphics[keepaspectratio, width = 0.45\linewidth]{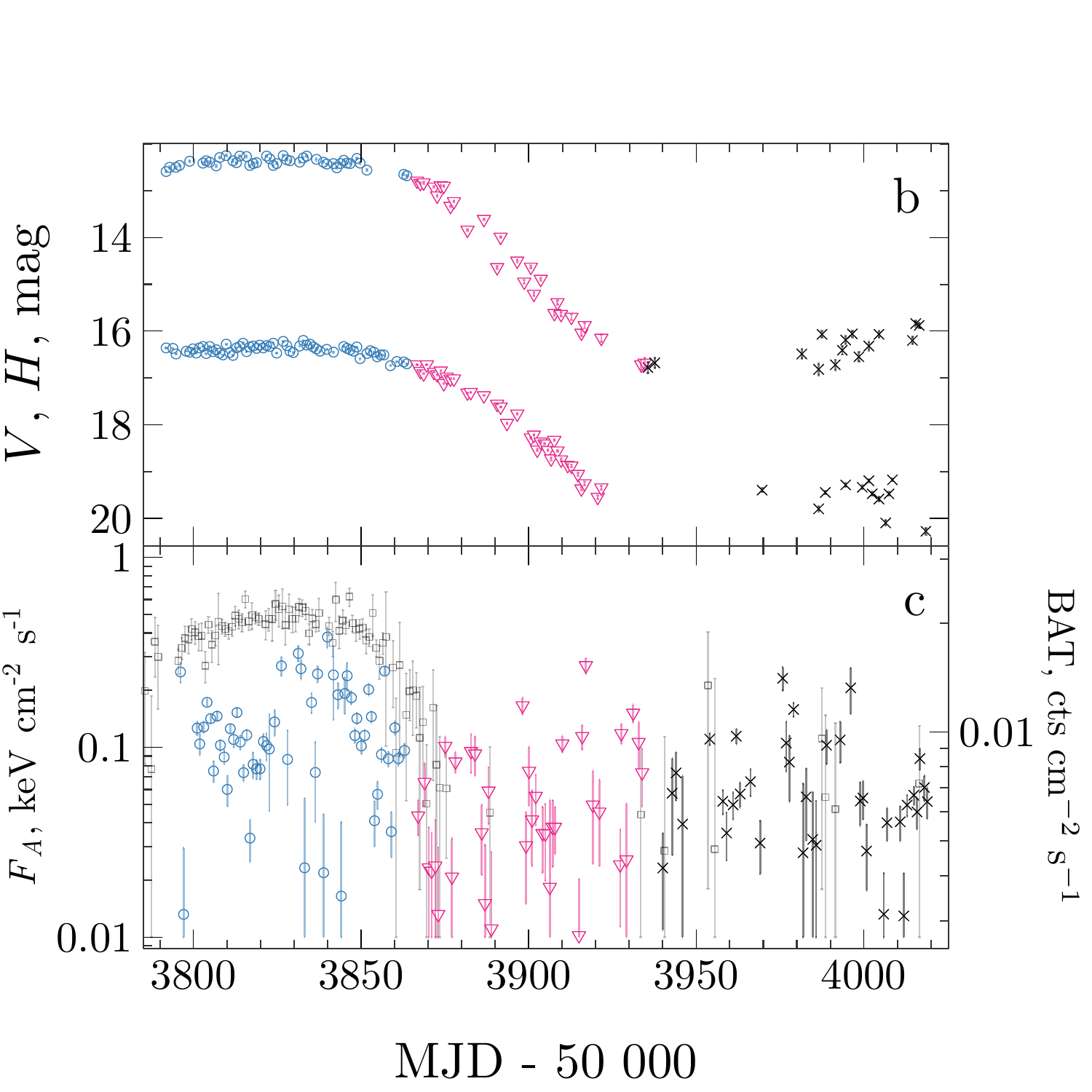}
   \caption{Same as Fig.~\ref{fig:CMD_LC_02}, but for the 2006 failed outburst.
    We use blue circles to mark the bright phase of the outburst and black crosses to mark the quiescent state.
     The transition is marked with pink triangles.}
    \label{fig:CMD_LC_F_01}
\end{figure*}

\begin{figure*}
\centering
       \includegraphics[keepaspectratio, width = 0.45\linewidth]{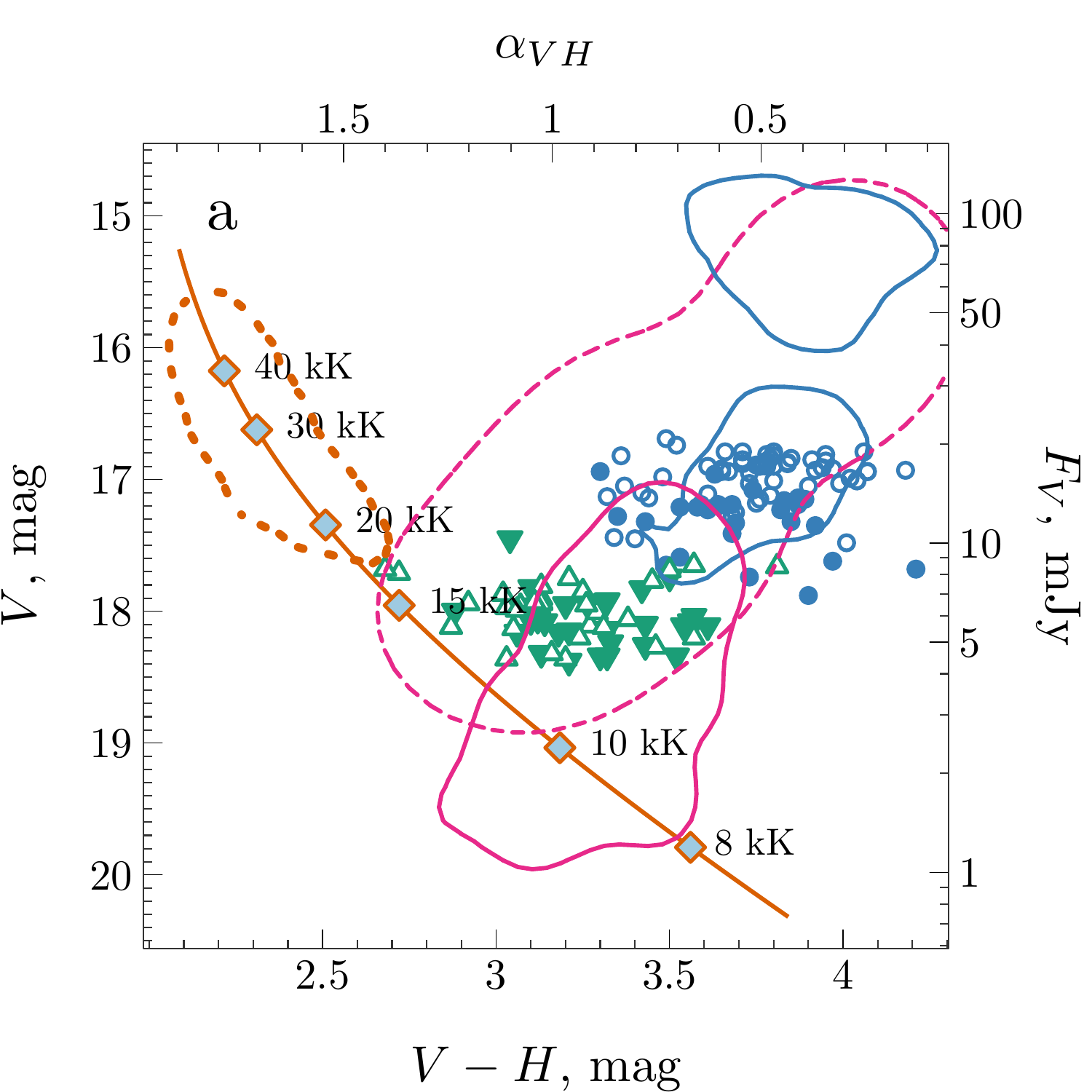}
       \includegraphics[keepaspectratio, width = 0.45\linewidth]{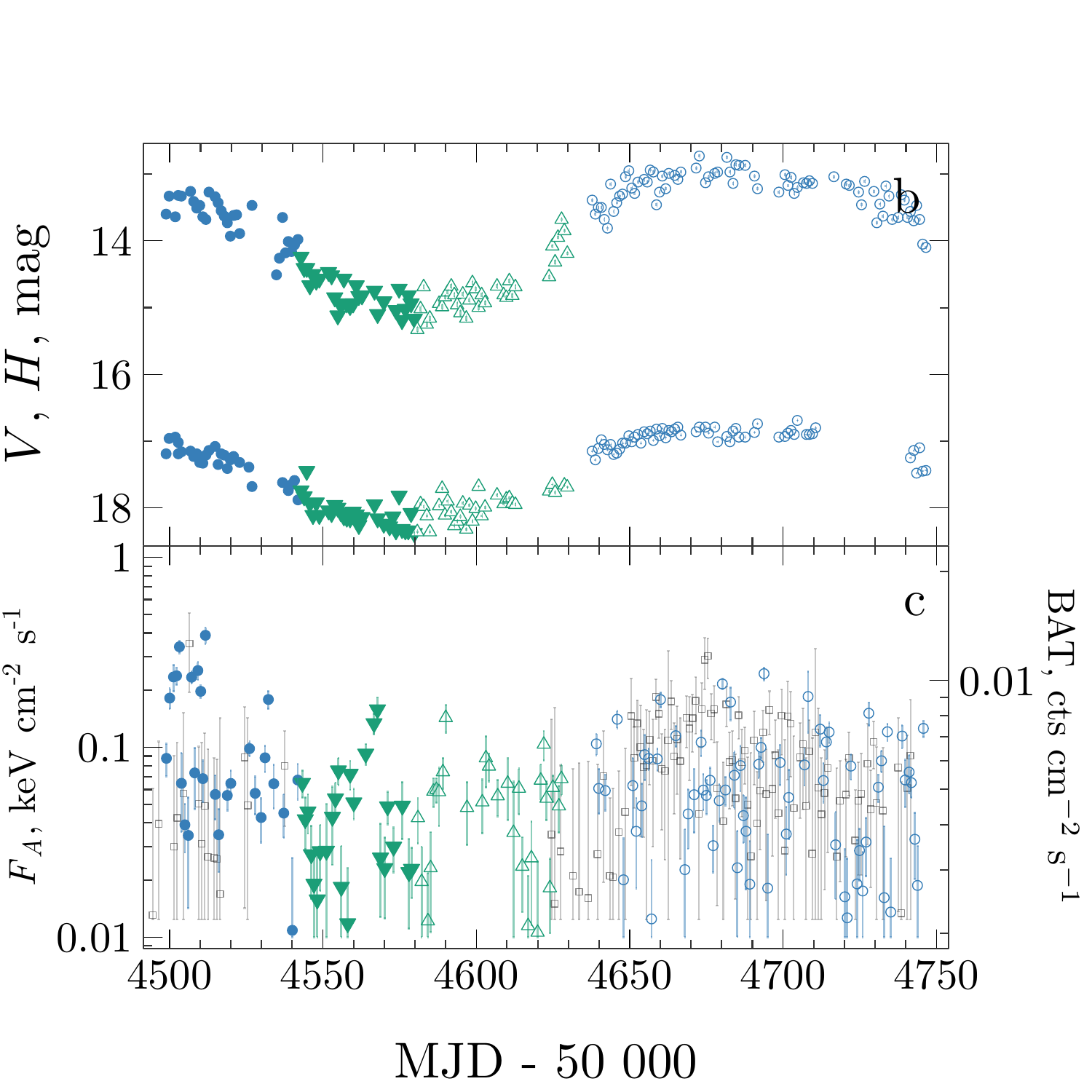}
   \caption{Same as Fig.~\ref{fig:CMD_LC_F_01}, but for the 2008 failed outburst. 
    We use filled green triangles to mark the transition from the initial bright phase to the faint phase and open green triangles to mark the rebrightening.} 
    \label{fig:CMD_LC_F_03}
\end{figure*}

\begin{figure*}
\centering
       \includegraphics[keepaspectratio, width = 0.45\linewidth]{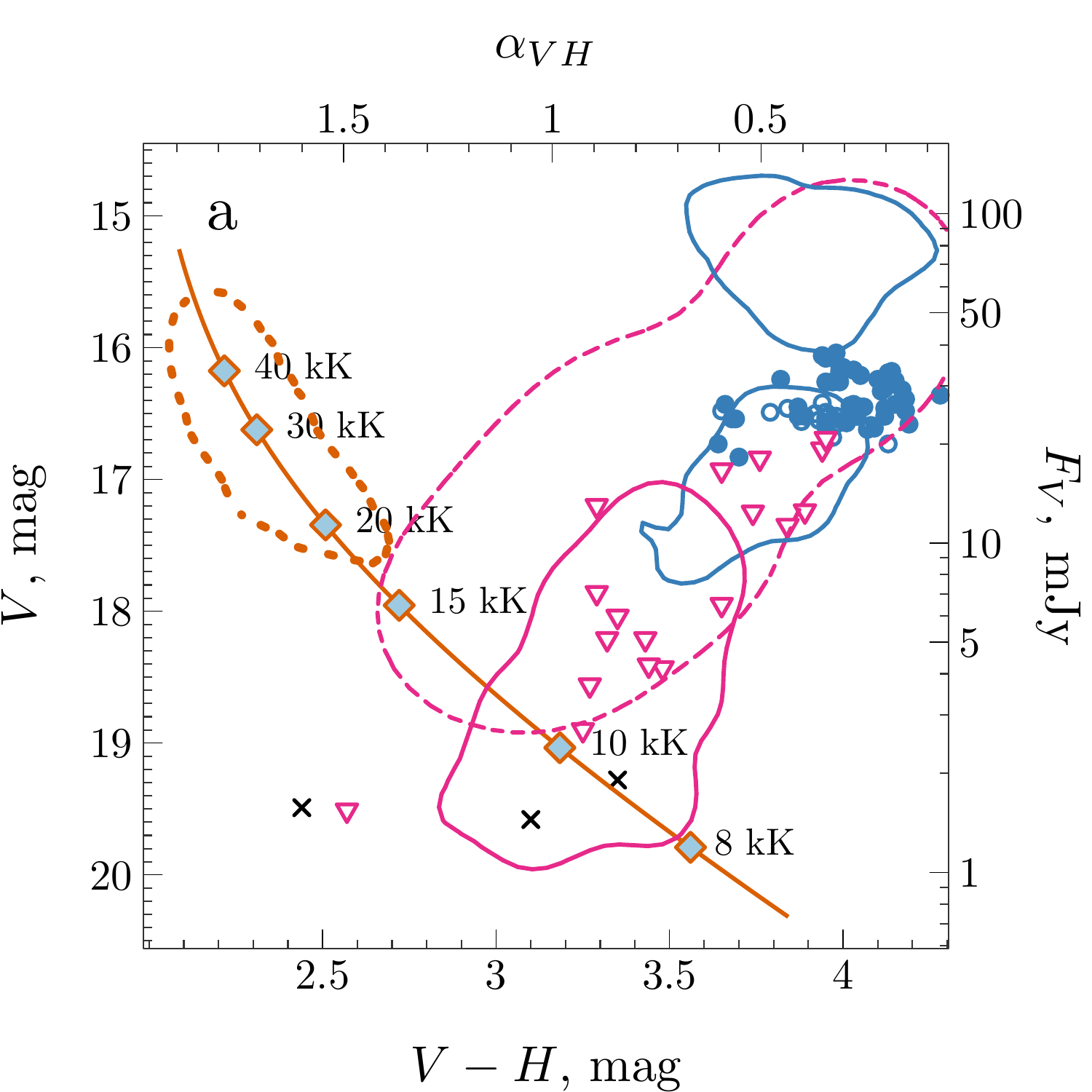}
      \includegraphics[keepaspectratio, width = 0.45\linewidth]{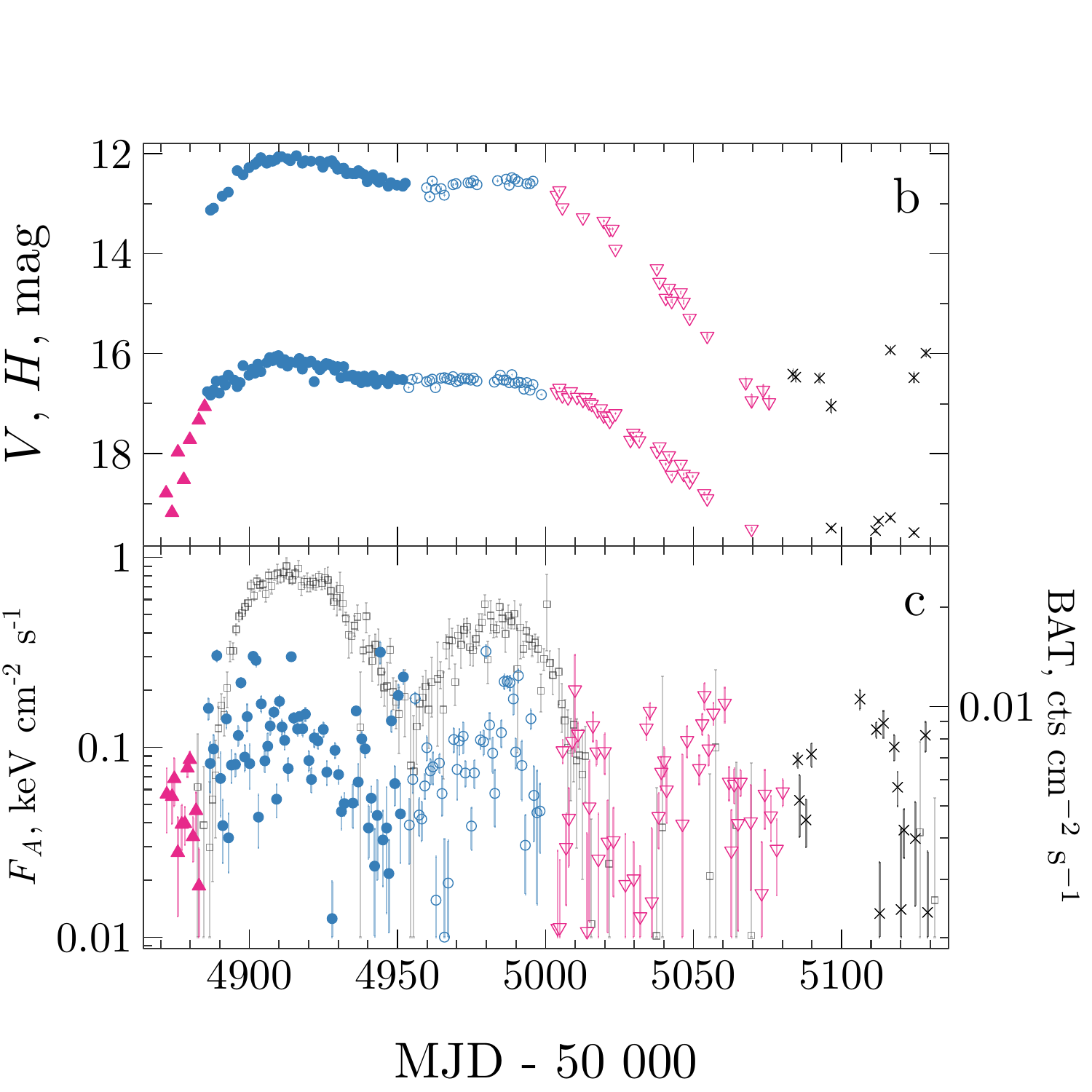}
   \caption{Same as Fig.~\ref{fig:CMD_LC_F_01}, but for the 2009 failed outburst.}
    \label{fig:CMD_LC_F_02}
\end{figure*}

The 2007 outburst demonstrates unusually short SS and a long DHS.
Both ASM and ONIR fluxes decayed after the HtS transition (Fig.~\ref{fig:CMD_LC_03}b,c), and the decay took about 100~d, as compared to typical 250--300~d for other outbursts. 
On the contrary, the DHS lasted over 100~d, in contrast to typical $\sim$50~d for other outbursts. 
It is even possible that this phase continues throughout the first half of 2008.

The ONIR light-curves of the 2010--2011 outburst show peculiar trend discontinuity during the SS (around \MJD{55360}, see Fig.~\ref{fig:CMD_LC_04}b).
Nevertheless, all SS data points align well with the blackbody model on the CMD (Fig.~\ref{fig:CMD_LC_04}a), hence, the observed break is caused by the sudden decrease of the color disk temperature.

\subsection{Failed outbursts}
\label{sec:failed}

We have three failed outbursts in 2002--2011, in 2006 (see Figs.~\ref{fig:CMD_LC_F_01}), in 2008 (see Fig.~\ref{fig:CMD_LC_F_03}) and in 2009 (see  Fig.~\ref{fig:CMD_LC_F_02}), when \GX~does not enter the SS, in contrast to regular outbursts.
The ONIR peak fluxes are a few times smaller than the RHS of regular outbursts, however, the colors are the same.
The duration of different phases of failed outbursts is comparable to those of regular outbursts. 

The 2006 and 2009 failed outbursts have similar duration and peak magnitudes.  
The light curves of the 2009 failed outburst have  double-peaked profile (both in the ONIR and the X-rays) and the $V-H$ colors of both peaks are the same. 
Interestingly though, the brightest episode of the 2009 outburst resembles the RHS of the 2004--2005 regular one in terms of phase duration, typical ONIR magnitudes and colors (compare Fig.~\ref{fig:CMD_LC_02}b and Fig.~\ref{fig:CMD_LC_F_02}b, blue dots).
This similarity makes it harder to distinguish between regular and failed outbursts at early phases.

The 2008 failed outburst peak is about 0.5 magnitude fainter than that of the other failed outbursts.
It consists of two peaks \citep{ATel1586, ATel1588}, both of them having colors and magnitudes typical to the DHS (see Fig.~\ref{fig:CMD_LC_F_03}a).
The second brightening is, on the other hand, considerably longer than the typical DHS. 
Between the two peaks the source dims and becomes bluer as if it were going to the SS, similar to the HtS transition.  
However, it reddens again and returns back to the DHS along a similar track on the CMD. 
Therefore, we color these phases on the CMD as HtS and StH. 
We note that the timescales are somewhat larger (40--50~d as opposed to 10--20~d; see Fig.~\ref{fig:CMD_LC_F_03}b, Table~\ref{tbl:intrvl}).
Though we consider the 2008 outburst as a separate one, there is a possibility that it can be a continuation of the exceptionally long 2007 outburst.

\begin{table}
    \centering
    \caption{Best-fit parameters for model (\ref{eq:model}) to the light curves in the soft state. Errors are 1$\sigma$.}
    \label{tbl:MDL_FITS}
 \begin{tabular}{cccr@{$~\pm~$}lc}
    \hline
    \hline
	                 Filter  &  $t^0$ & $m^0_j$ &  \multicolumn{2}{c}{$\mu_j$}        & $\epsilon_j$ \\
	                         &   (day)    &  (mag)    &  \multicolumn{2}{c}{(mmag~day$^{-1}$)} & (mag)          \\
  	\hline
  	    \noalign{\vskip 1ex}
  	    \multicolumn{5}{c}{ Decaying phase of the 2002--2003 outburst}\\[1ex]
  	      $ V $ &  \multirow{4}{*}{52742} &  $ 17.50 \pm 0.06 $ &  9 & 2  &       $ 0.16 \pm 0.02 $ \\ 
          $ I $ &   &      $ 16.29 \pm 0.05 $ &  8 & 2  &       $ 0.14 \pm 0.02 $ \\ 
          $ J $ &   &      $ 15.32 \pm 0.09 $ &  6 & 3  &       $ 0.20 \pm 0.03 $ \\ 
          $ H $ &   &      $ 14.80 \pm 0.06 $ &  3 & 2  &       $ 0.13 \pm 0.02 $ \\[1ex]
	     \multicolumn{5}{c}{Rising phase of the 2004--2005 outburst}  \\[1ex]
          $ V $ &  \multirow{4}{*}{53229} &      $ 16.48 \pm 0.03 $ &     $-$4 & 1  &       $ 0.11 \pm 0.01 $ \\ 
          $ I $ &   &      $ 15.46 \pm 0.03 $ &     $-$4 & 1  &       $ 0.11 \pm 0.01 $ \\ 
          $ J $ &   &      $ 14.43 \pm 0.07 $ &    $-$2 & 1  &       $ 0.17 \pm 0.03 $ \\ 
          $ H $ &   &      $ 14.17 \pm 0.06 $ &     $-$3 & 1  &       $ 0.11 \pm 0.02 $ \\[1ex]
  	    \multicolumn{5}{c}{Decaying phase of the 2004--2005 outburst} \\[1ex]
          $ V $ &  \multirow{4}{*}{53477} &      $ 17.35 \pm 0.06 $ &  15 & 2  &       $ 0.20 \pm 0.02 $ \\ 
          $ I $ &   &      $ 16.18 \pm 0.05 $ &  13 & 2  &       $ 0.17 \pm 0.02 $ \\ 
          $ J $ &   &      $ 15.15 \pm 0.04 $ &  11 & 1  &       $ 0.14 \pm 0.02 $ \\ 
          $ H $ &   &      $ 14.69 \pm 0.04 $ &  8 & 1   &       $ 0.12 \pm 0.01 $ \\[1ex]
    	 \multicolumn{5}{c}{Decaying phase of the 2007 outburst} \\[1ex]
          $ V $ &  \multirow{4}{*}{54240} &      $ 17.51 \pm 0.06 $ &  12 & 2  &       $ 0.15 \pm 0.02 $ \\ 
          $ I $ &   &      $ 16.36 \pm 0.06 $ &  11 & 2  &       $ 0.16 \pm 0.02 $ \\ 
          $ J $ &   &      $ 15.28 \pm 0.06 $ &  6 & 2  &       $ 0.14 \pm 0.02 $ \\ 
          $ H $ &   &      $ 14.78 \pm 0.09 $ &  2 & 3   &       $ 0.18 \pm 0.03 $ \\[1ex] 
  	    \multicolumn{5}{c}{Rising phase of the 2010--2011 outburst} \\[1ex]
          $ V $ &  \multirow{4}{*}{55302} &      $ 16.53 \pm 0.04 $ &     $-$9 & 1  &       $ 0.16 \pm 0.01 $ \\ 
          $ I $ &   &      $ 15.50 \pm 0.03 $ &     $-$10 & 1  &       $ 0.15 \pm 0.01 $ \\ 
          $ J $ &   &      $ 14.57 \pm 0.04 $ &     $-$8 & 1  &       $ 0.14 \pm 0.01 $ \\ 
          $ H $ &   &      $ 14.20 \pm 0.04 $ &     $-$7 & 1  &       $ 0.15 \pm 0.01 $ \\[1ex] 
	    \multicolumn{5}{c}{Decaying phase of the 2010--2011 outburst} \\[1ex]
          $ V $ &  \multirow{4}{*}{55606} &      $ 17.91 \pm 0.14 $ &  45 & 10  &       $ 0.21 \pm 0.05 $ \\ 
          $ I $ &  &      $ 16.68 \pm 0.11 $ &  38 & 8  &       $ 0.19 \pm 0.04 $ \\ 
          $ J $ &  &      $ 15.57 \pm 0.11 $ &  32 & 7  &       $ 0.18 \pm 0.04 $ \\ 
          $ H $ &  &      $ 15.05 \pm 0.16 $ &  22 & 11  &       $ 0.22 \pm 0.06 $ \\ 
	\hline
\end{tabular}
\end{table}

We note that the distinction between the failed and regular outbursts is not possible from the ONIR data alone at their earlier stages. 
However, it might be possible using the peak $V$-magnitude. 
If the source is brighter that $V=16$\,mag during the RHS, it will transit to the SS.  
On the other hand, if it is fainter, the outburst can be either regular (2004--2005 outburst, see Fig.~\ref{fig:CMD_LC_02}) or failed (2009 outburst, see Fig.~\ref{fig:CMD_LC_F_02}).

\subsection{Evolution of the non-thermal component}\label{sec:fitting}

It is known that the changes of ONIR colors during the HtS and StH transitions are related to quenching/recovery of the red, non-thermal component \citep[see e.g.][]{Fender1999, JBOM01, Corbel2002}. 
To understand the origin of the non-thermal emission, it is important to obtain a reliable estimate of its spectral shape by subtracting the contribution of the thermal (disk) component. 
The task is, however, complicated by the fact that the spectrum of the thermal component (e.g. disk temperature) is an unknown function of time.
The spectral index of the non-thermal component may allow us to distinguish between two alternatives: the jet \citep{Fender2001, Gallo2007, Uttley2014} and the hot flow \citep{Veledina2013, Poutanen2014a,Kajava2016}.
The contribution of the disk can be estimated by fitting the SS light-curves in different filters, and extrapolating the obtained fit to the transitions and the hard state \citep[see, e.g.,][]{Buxton2012, Dincer2012,Poutanen2014}.
The weakness of this approach is related to freedom in the choice of the fitting function and the fitting interval, which leads to an uncertainty in the spectrum of the non-thermal component.

In general, the emission of the irradiated accretion disk is expected to follow the fast rise -- exponential decay profile \citep[and a constant level of emission in quiescence, see e.g.][]{Chen1997}.
The exponential decay profile was fit to the 2000 outburst data of XTE~J1550--564 \citep{Poutanen2014}, while \GX~shows both episodes of decay and rise during the SS. 
Therefore, we fit parts of the SS light curves using the following model:
\begin{equation}
    \label{eq:model}
    m_j(t) = m_j^0 + \mu_j \left(t - t^0\right) + \epsilon_j,
\end{equation}
where $\mu_j$ is the slope, which can be positive or negative. 
Here $m_j$ is the observed magnitude in $j$-th ONIR filter, $m_j^0$ is the constant, $t$ is the time and $t^0$ is either the beginning (if the fitted data set covers the first part of the SS) or the end time of the fitted data set (if it covers the last part of the SS), and $\epsilon_j$ is the intrinsic scatter, which is partially caused by superhump variability \citep{Dincer2012, Kosenkov2018}.
We use Bayesian inference to estimate these parameters and list them in Table~\ref{tbl:MDL_FITS}.

Once the fit to the SS is obtained, we extrapolate the model for 30 days from the start or the end of the SS, which usually cover the transition phase and a part of the hard state.
The errors on the fluxes of the non-thermal component are mostly coming from the uncertainties in the parameters of the fitted model and they increase towards the hard state.
On the other hand, the contribution of the disk also drops, and it falls down to about 5\% in the RHS.

\begin{figure*}[h!]
    \centering
    \includegraphics[keepaspectratio, width = 0.9\linewidth]{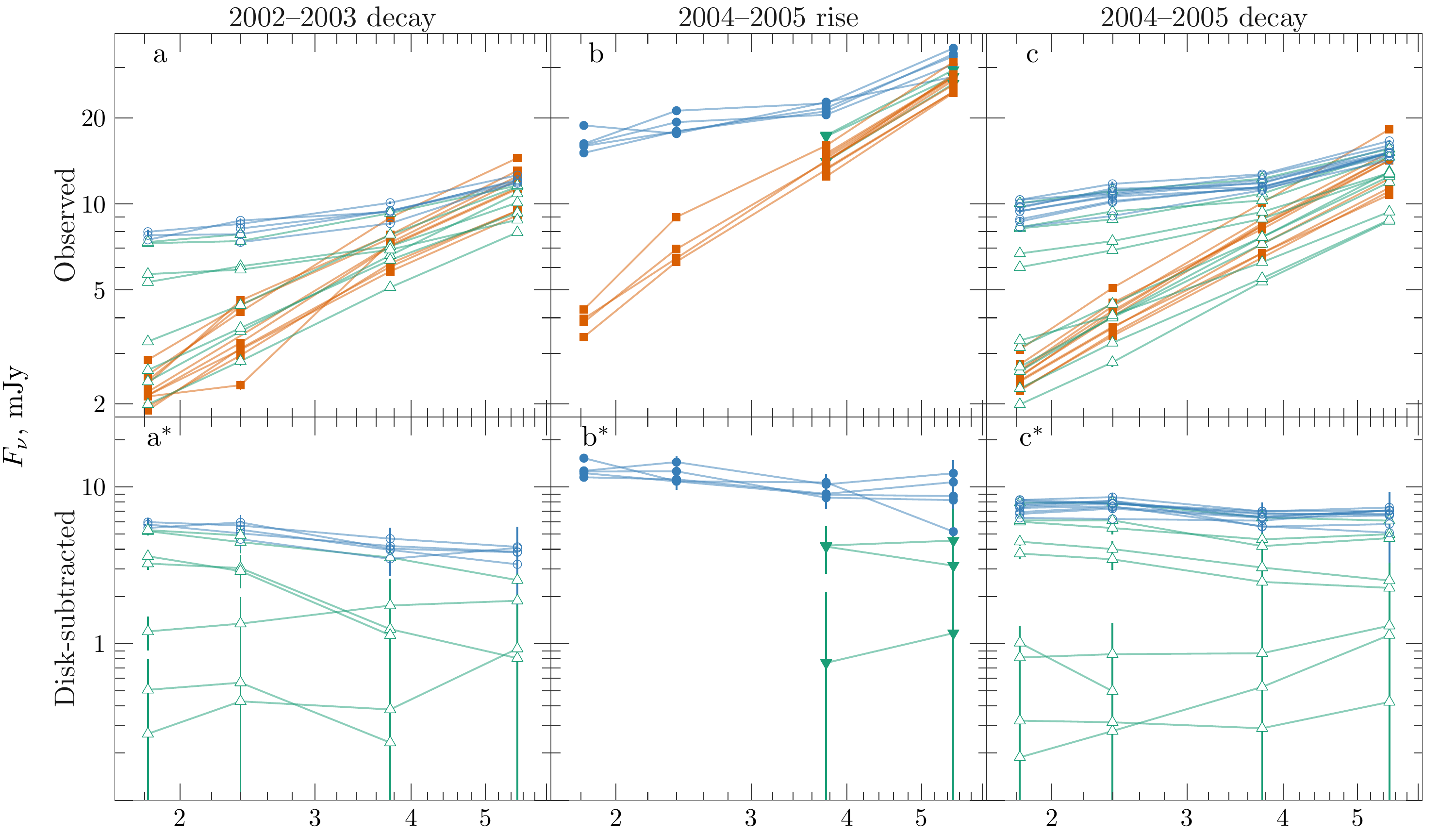}
    \includegraphics[keepaspectratio, width = 0.9\linewidth]{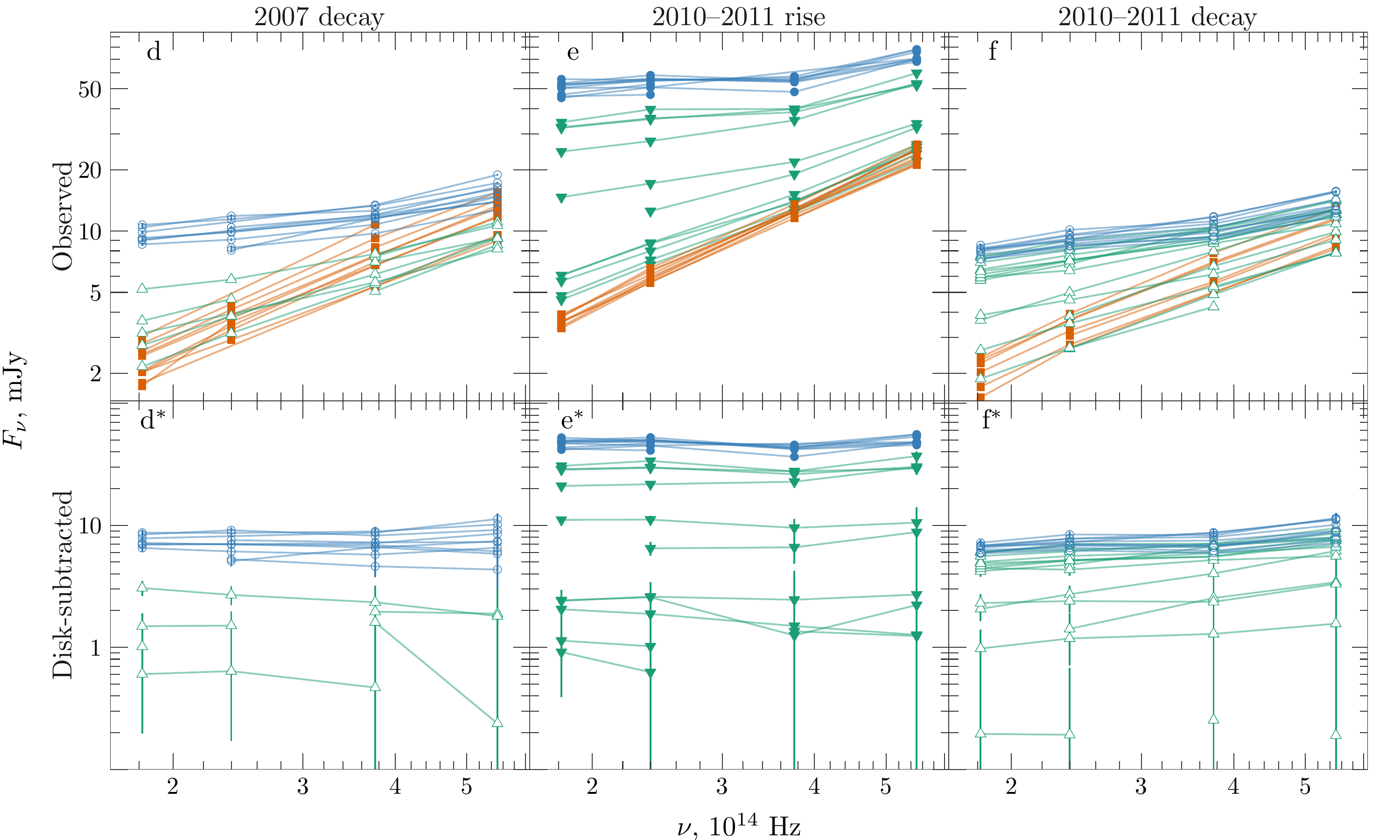}
    \caption{Sample of spectra from the soft state (orange squares),  transition (green triangles) and the hard-state spectra (blue circles). 
    Open symbols are for the soft to hard state transitions, while the filled symbols are for the hard to soft state transitions. 
    Upper panels (a-f) show total observed spectra of the decaying part of the 2002--2003 outburst, rising and decaying parts of the 2004--2005 outburst, the 2007 outburst and the rising and decaying parts of the 2010--2011 outburst, respectively.
    Lower panels (a$^*$-f$^*$) show corresponding spectra after subtraction of the thermal component.
    Fluxes are corrected for the interstellar extinction, errors are 1$\sigma$.}
    \label{fig:SPEC_1_4}
\end{figure*}

The fitted models are shown in Figs.~\ref{fig:CMD_LC_01}-\ref{fig:CMD_LC_04}(b) with the solid black lines. 
The gray areas around the model denote 1$\sigma$ uncertainties. 
The model curve spans over the SS data used for fitting (orange squares), the transition phase (if present, green triangles) up to the nearest hard state (blue circles).
The total spectra and the spectra of the non-thermal component are shown in Fig.\,\ref{fig:SPEC_1_4}.
The fluxes are corrected for the interstellar extinction assuming $A_V = 3.58$\,mag (see Sect.~\ref{sec:extinction}).
 
In the majority of cases, both the hard state and the state transition spectra of the non-thermal component appear to be nearly flat ($\alpha_{VH}^\mrm{nth} = 0.06 \pm 0.12$, the best examples are the 2004--2005 DHS  and 2010--2011 RHS, see  Fig.~\ref{fig:SPEC_1_4}c$^*$,e$^*$). 
During the transitions, the spectral shape of the non-thermal component is stable.
We also note that there is no apparent difference between spectral shapes of that component in the StH and HtS transitions.

The spectral slope $\alpha_\mrm{ONIR}$ is affected by the assumed value of $A_V$. The impact of $A_V$ can be estimated using Eq.\,(\ref{eq:extin_coeff_short}).
For a relatively low value of $A_V \approx 3.0$\,mag \citep{Kong00, HBMB05}, $\alpha_{VH}^\mrm{nth}=-0.34$, while for a relatively high $A_V = 3.7$\,mag \citep{Zdziarski1998, Buxton2012} we get $\alpha_{VH}^\mrm{nth} = 0.15$.
The changes in $A_V$ are also reflected in the curvature of the blackbody track on the CMD located within the SS contour 
(see discussion in Section~\ref{sec:evolution_template}).

\begin{figure*}
    \centering\includegraphics[keepaspectratio, width = 0.95\linewidth]{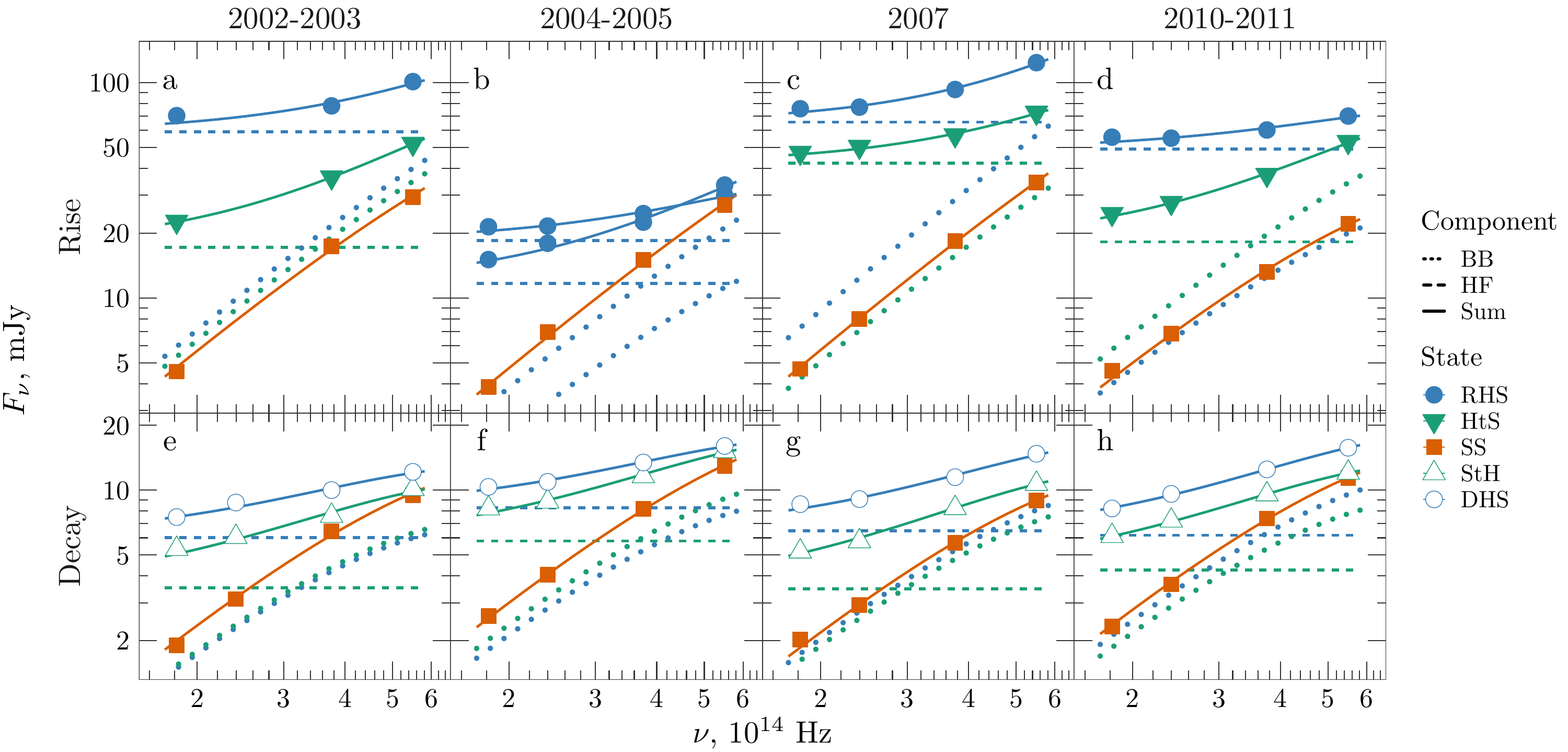}
    \caption{
       Selected spectra of the rising (upper panels) and decaying (lower panels) stages of regular outbursts.
       1$\sigma$ errors are comparable to the symbol size.
        Solid orange and dotted blue and green lines correspond to the blackbody (BB) component in the soft and hard states and transitions, respectively.  
        Dashed blue and green lines correspond to the power-law spectra of the hot flow (HF). 
        Solid lines give the sums of the corresponding blackbody and the hot flow components. 
        In panel (b) we do not plot HtS spectrum because there are no observations with simultaneous measurements in three or more ONIR filters.
        Instead, we show two RHS spectra, separated by $\sim 150$~d, to highlight the evolution of both non-thermal and disk components (see also Fig.~\ref{fig:CMD_LC_02}).
    }
    \label{fig:spec_idv}
\end{figure*}

To illustrate the results of our spectral decomposition, we select a sample of the RHS, HtS, SS, StH and DHS data of each regular outburst.
We fit a model consisting of a power law of index $\alpha=0$ to account for the non-thermal component (except for the SS) and a blackbody to the data corrected for interstellar extinction. 
The blackbody normalization is allowed to vary between the outbursts.
The results are shown in Fig.~\ref{fig:spec_idv}.
This simple model reproduces the data very well.

\subsection{Broad-band spectra in the hard state}
\label{sec:IR}

\begin{figure*}
    \centering
    \includegraphics[keepaspectratio, width=0.9\linewidth]{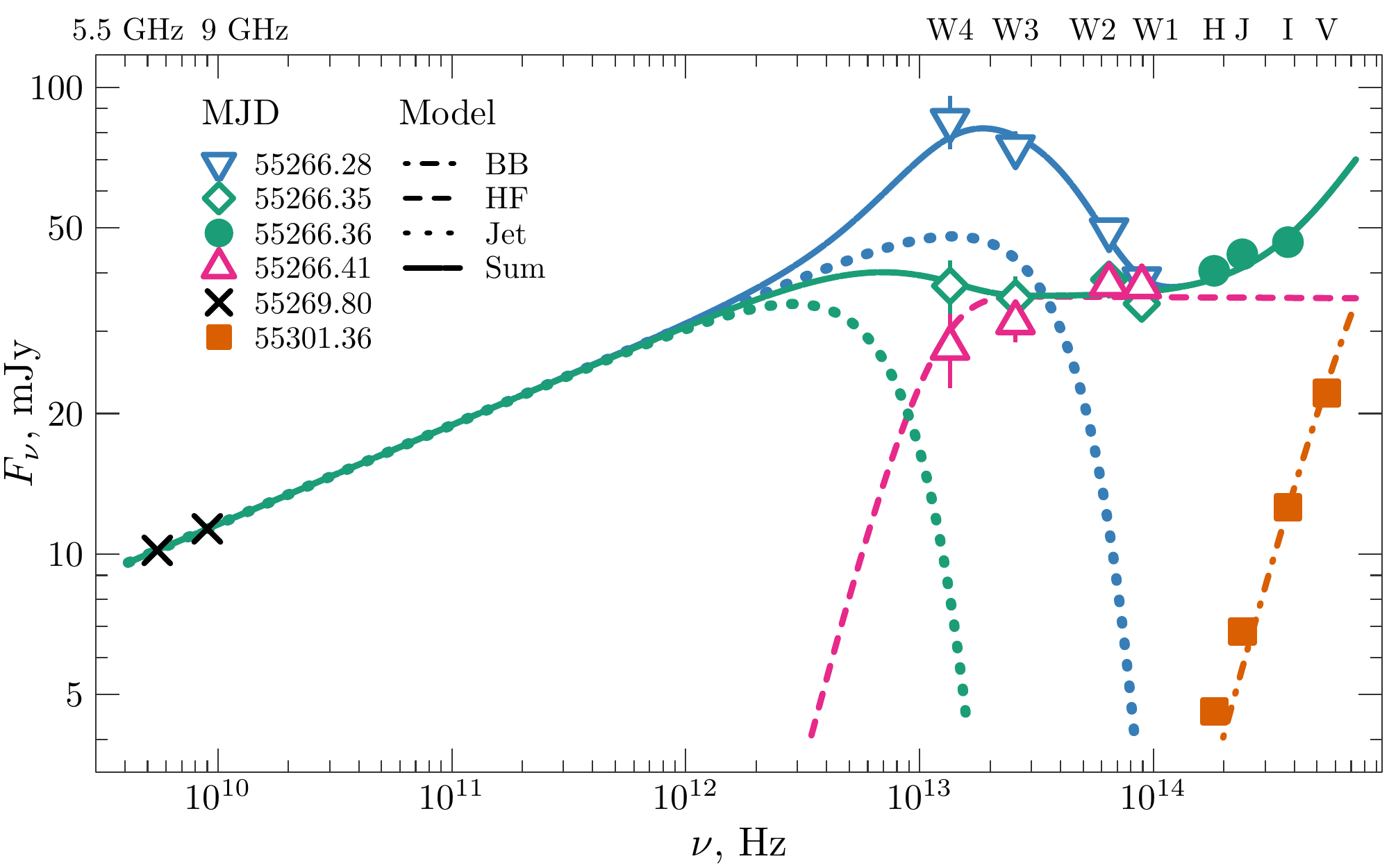}
    \caption{
        Broad-band spectra of \GX\ using the radio ATCA, mid-IR \textit{WISE} and  ONIR SMARTS data.  
        The figure shows four distinct spectral shapes for the source.
        Orange squares show the soft-state data from \MJD{55301.36}, the filled green circles and open green diamonds are the quasi-simultaneous ONIR and mid-IR data on \MJD{55266.35}. 
        Open blue downward-facing triangles and open pink upward facing triangles show the mid-IR spectra measured 1.7\,h before and after the quasi-simultaneous observation, respectively. 
        Black crosses are one of the closest available ATCA radio observation (\MJD{55269.80}) that is used to constrain the jet properties. 
        The lines show different model components:  dot-dashed orange line is the blackbody component, dashed pink line is for the hot-flow component, the dotted lines correspond to the jet model with different cutoff frequencies, and solid lines give the sum of the three (blackbody + hot flow + jet) components. 
        The sum of the blackbody and hot flow components for \MJD{55266.41} is not shown for convenience, because it fully coincides with the hot flow curve in the mid-IR.
        1$\sigma$ errors are comparable with the symbol size. 
        The \textit{WISE} fluxes are adjusted by the normalization factors inferred from the fit (see Table~\ref{tbl:mdl_fit_param}).  
    }
    \label{fig:broadband}
\end{figure*}

The hard-state mid-IR spectra of \GX~can help us identify the nature of the non-thermal component. 
\GX~was observed in the mid-IR with \textit{WISE} on \MJD{55265.88--55266.88}  \citep{Gandhi2011}. 
The 13 mid-IR spectra can be divided into two groups.
At lower fluxes, the \textit{WISE} spectra are nearly flat (see  spectra \#1, 2, 7, and 8 in their fig.~3) and    lie on the continuation of the ONIR non-thermal component.  
At higher fluxes, the spectra are much softer and most of them have a large curvature (e.g. spectra \#3--6, 9, 12--13; see also blue triangles in  Fig.~\ref{fig:broadband}). 

The source shows strong, by a factor of three, variability in all four \textit{WISE} filters on the timescales of hours \citep[see fig.~2a and table~1 in][]{Gandhi2011}. 
Variability amplitude increases towards longer wavelengths. 
Extrapolation of the \textit{WISE} spectra to the ONIR bands predict an order of magnitude variation there. 
However, the observed ONIR non-thermal component is nearly constant. 
This suggests that there are two non-thermal components: one highly variable dominating during high mid-IR flux episodes and another, more stable, emitting in a broad range from mid-IR to ONIR. 
There is a caveat though that the typical \textit{WISE} exposure is about 10~s, which is much smaller than the typical 300~s exposure in the ONIR bands.

To understand the nature of these components, we try to construct the source broad-band spectra.  
All available 13 \textit{WISE} observations occurred during 24 hours on \MJD{55265.9--55266.9}. 
There is one observation on \MJD{55266.3481}, which is just 16 min before the SMARTS observations on \MJD{55266.3594}. 
In addition we also choose two nearest \textit{WISE} observations 1.7\,h before and after the quasi-simultaneous one, on \MJD{55266.28} and on 55266.41. 
In order to estimate contribution from the disk blackbody, we use the SS ONIR observation on \MJD{55301.36}, because after the HtS transition the ONIR flux was nearly constant (see Fig.\,\ref{fig:CMD_LC_04}).
Furthermore, we select the closest in time ATCA 5.5 and 9~GHz measurements obtained on \MJD{55269.80} \citep{Corbel2013}. 
Albeit the radio and mid-IR data are taken on different dates, the extrapolation of the hard radio spectrum matches the high-flux \textit{WISE} data. 
The connection between the radio and mid-IR fluxes and their high variability on short timescales are hallmarks of the jet emission as seen in blazars \citep{Tavecchio17,Zacharias18}. 
A simple jet model of \citet{Blandford1979}, which was often invoked to explain the radio to NIR spectra of \GX\ \citep[see e.g.][]{Corbel2002,Gandhi2011,Rahoui2012,Corbel2013a}, does not address the presence of the sharp cutoff in the mid-IR spectra and the dichotomy of mid-IR spectral slopes.

We suggest that the broad-band spectrum of \GX\ consists of (at least) three components. 
We model the disk component by the blackbody.
We use a power law with spectral index close to zero, which  transforms to Rayleigh-Jeans spectrum at lower-frequencies, to model the hot flow.
The jet is described by a power law with a high-frequency cutoff. 
The model can be described by the following equations:
\begin{equation}
    \begin{aligned}
        F_\nu^\mrm{bb} &= N_{\rm bb}\ B_\nu(T),\\
        F_\nu^\mrm{hf} &= N_{\rm hf}  \left(\frac{\nu}{10^{14}\,{\rm Hz}}\right) ^ \alpha 
        \left\{ 1 - \exp\left[ - \left( \frac{\nu}{\nu_{\rm hf}} \right)^{2-\alpha}  \right] \right\} , \\
        F_\nu^\mrm{jet} &= N_{\rm jet} \left(\frac{\nu}{10^{10}\,{\rm Hz}}\right) ^ \beta \exp\left[ -\left(\frac{\nu}{\nu_{\rm jet}}\right)^\gamma\right],\\
        F_\nu^\mrm{tot} &= F_\nu^\mrm{bb} + F_\nu^\mrm{hf} + F_\nu^\mrm{jet},
    \end{aligned}
    \label{eq:spec_dec}
\end{equation}
where $F_\nu^\mrm{bb}$,~$F_\nu^\mrm{hf}$,~$F_\nu^\mrm{jet}$ and $F_\nu^\mrm{tot}$ are the model fluxes of the disk, hot flow, jet components and their sum, respectively, $N_{\rm bb}$, $N_{\rm hf}$ and $N_{\rm jet}$ are the model normalizations,  $T$ is the blackbody temperature, $\nu_{\rm hf}$ is the low-frequency cutoff of the hot flow component, $\nu_{\rm jet}$ is the high-frequency cutoff of the jet spectrum, $\alpha$ and $\beta$ are the corresponding power-law slopes, and $\gamma$ is the index of the super-exponential cutoff. 
The value of $\log N_{\rm bb}$ is fixed at 4.09 derived in Sect.\,\ref{sec:extinction}.

\begin{table*}
    \centering
 \caption{Best-fit parameters for model given by Eq.~(\ref{eq:spec_dec}) for the hard-state radio to ONIR spectra and the soft-state ONIR spectrum of \GX\ shown in Fig.\,\ref{fig:broadband}.}
    \label{tbl:mdl_fit_param}
    \begin{tabular}{c|c|ccc|cccc|c} 
        \hline
        \hline
        & Blackbody\tablefootmark{a}  & \multicolumn{3}{c|}{Hot flow\tablefootmark{b}} & \multicolumn{4}{c|}{Jet\tablefootmark{c}} &\\
        \hline
        Dates & $T$ & $N_{\rm hf}$ & $\nu_{\rm hf}$ & $\alpha$ &  $N_{\rm jet}$ & $\nu_{\rm jet}$ &     $\beta$ & $\gamma$ & $N_{WISE}$\tablefootmark{d}\\
        (MJD) & ($10^4$~K) & (mJy) & ($10^{13}$~Hz) & &  (mJy) & ($10^{13}$~Hz) &  & & \\
        \hline
        55266.28 & \multirow{4}{*}{$3.10 \pm 0.02$} & \multirow{3}{*}{$35.5 \pm 0.2$} &            \multirow{3}{*}{$1.1 \pm 0.4$}& \multirow{3}{*}{$-0.03 \pm 0.02$} & \multirow{3}{*}{11.6 $\pm$ 0.1} & 4.6 $\pm$ 0.7  & \multirow{3}{*}{0.22 $\pm$ 0.03} &  \multirow{3}{*}{$1.8 \pm 0.4$} & 0.97 $\pm$ 0.06\\
        55266.35  & &   &   &    &  & \phantom{4}0.97 $\pm$ 0.25 & &  & 0.74 $\pm$ 0.02 \\ 
        55266.41  & &   &   &     &   & $<0.73$ &  &  & 0.76 $\pm$ 0.02 \\ 
        55301.36  & & ...  & ... &  ...   &  ... & ... & ... &  ... & ... \\ 
        \hline
    \end{tabular}
   \tablefoot{
   \tablefoottext{a}{Blackbody normalization was fixed at $\log N_{\rm bb} = 4.09$ and blackbody parameters are the same for all four spectra.}   
   \tablefoottext{b}{Hot flow parameters are the same for the three hard-state spectra.}  
   \tablefoottext{c}{Jet parameters, except cutoff frequency, are the same for the three hard-state spectra.}
   \tablefoottext{d}{Cross-calibration factor of the \textit{WISE} data.}
   Errors are 1$\sigma$.
   }
\end{table*}

Fig.~\ref{fig:broadband} shows the RHS spectra of \GX\ as observed by ATCA, \textit{WISE} and {SMARTS} together with one SS ONIR spectrum as well as the best-fit models. 
We assume that the hot flow spectrum is the same for all three mid-IR spectra. 
For the jet component, we use different cutoff frequencies, because normalization and power-law slope are determined by the shape of the radio data, which is the same in all cases. 
Due to the difference in the exposure times of \textit{WISE} ($<$10\,s) and SMARTS (300\,s) as well as some systematic uncertainties in absolute flux of \textit{WISE}, we apply additional cross-calibration factor to the mid-IR fluxes, which is treated as a free parameter. 
The best-fit parameters are given in Table~\ref{tbl:mdl_fit_param}.

We see that on \MJD{55266.28} the jet outshines the hot flow by a factor of 2--3 in $W4$ and $W3$ filters. 
The overall, radio to ONIR, spectrum cannot be described by a broken power law plus a blackbody. 
This implies that the spectrum above $\sim 3\times 10^{13}$\,Hz is not consistent with an optically-thin, power-law-like synchrotron spectrum predicted by a standard jet model of \citet{Blandford1979}. 
The sharp cutoff below $10^{14}$\,Hz cannot be described even by a simple exponential cutoff but is well fit only by a super-exponential with the index $\gamma\approx2$. 
If we interpret the power-law spectrum from radio to the mid-IR as a combination of self-absorbed synchrotron peaks, such a cutoff can only be obtained if the electron distribution producing the mid-IR emission has a very steep slope. 

On \MJD{55266.35} the jet likely contributes to the $W4$ filter, but the hot flow takes over at higher frequencies. 
The \textit{WISE}  spectrum obtained on \MJD{55266.41} is the hardest one with $\alpha=0.16$ having a clear curvature below $2\times 10^{13}$\,Hz, which can be associated with the synchrotron self-absorption frequency of the hot-flow spectrum (dashed pink line). 
Therefore  the hot flow likely dominates in all mid-IR filters and contributes to the ONIR flux together with blackbody component.
The measurement of the low-frequency cutoff allows us to constrain the size of the hot flow (corona) to be \citep[eq.\,13 in][]{Veledina2013} $R_{\rm hf}\approx 3\times 10^{9}\,(10^{13}\,\mbox{Hz}/\nu_{\rm hf}) \approx 2.7\times10^9$\,cm, which corresponds to 900 Schwarzschild radii for a 10 solar-mass black hole. 
For this observation, we obtain only an upper limit on the jet cutoff frequency, because the contribution of the jet to the mid-IR is negligible.

\section{Discussion}
\label{sec:discussion}

\subsection{ONIR spectral properties}
\label{sec:evolution_template}

The magnitude-magnitude diagrams \citep[see][]{Buxton2012} can be used to emphasize the difference between hard and soft states and highlight the state transition tracks. 
However, these diagrams do not provide sufficient information about the shape of the spectra in each of the observed states and make it harder to separate RHS and DHS data points as they roughly follow the same broken power-law track (together with the data from quiescence, see e.g. fig.~9 in \citealt{Buxton2012}). 
Alternatively, one can use color-magnitude diagrams \citep{Maitra2008,Russell2011,Poutanen2014}. 
CMD plots benefit from the fact that they highlight both changes in the observed fluxes and in the shape of the spectra.
The characteristic hysteresis pattern tracked by the source on the CMD can be related to the hysteresis observed in the X-ray data \citep{Poutanen2014}. 
In three out of four regular outbursts, \GX~was observed to have nearly identical HtS and StH transitions patterns on a CMD (see Sect.~\ref{sec:ONNIRvsXRAY} for a discussion; also fig.~1 in \citealt{Corbel2013}, fig.~1 in \citealt{MunozDarias2008}), while during the HtS transition of the rising phase of the 2002--2003 outburst the X-ray fluxes were a few times smaller than the typical values (Fig.~\ref{fig:XR}b,c). 
Interestingly, though the 2002--2003 HtS transition starts at lower ONIR fluxes, it arrives at the same SS flux with a color temperature of 30--35~kK.  

Recently, the CMD  was used to study the spectral evolution of the  black hole X-ray binary XTE~J1550--564 during its 2000 outburst  \citep{Poutanen2014},  which was very similar to that of \GX.  
Notably, the highest SS disk temperature of 16~kK reached by XTE~J1550--564 is substantially smaller than the temperature of 50~kK in \GX~(see Fig.~\ref{fig:CMD_LC_04}). 
The inferred SS temperatures can be affected if $A_V$ is overestimated. 
We note, however, that for a lower $A_V$ the disk temperature becomes smaller, leading to a different curvature of the blackbody track on the CMD, inconsistent with the observed shape drawn by the SS data points (see Fig.~\ref{fig:TEMPL}, solid orange line and dotted orange contour). 
Further, the UV observations of \GX~carried out during the early phase of the 2010--2011 SS suggest that the peak of the blackbody component lies at wavelengths shorter than $\approx 2000$~\AA, beyond the near-UV \citep[see fig.~7 in][]{CB11}, which imposes a lower limit of $\approx 15$~kK on the blackbody temperature at the beginning of the SS.
Unlike XTE~J1550--564, which shows a typical exponential decay of ONIR fluxes during its SS, ONIR fluxes of \GX~ increase shortly after the HtS transition in two out of four regular outbursts.
As a result, the highest color temperatures of \GX~are observed somewhere in the middle of the SS, while the highest blackbody temperature of XTE~J1550--564 is found at the beginning of the SS.

The disk temperatures depend on the flux irradiating the outer parts of the disk.
This flux is defined by the disk geometry, luminosity of the central X-ray source and its emission pattern, which in turn depends on the black hole spin.
The accretion disk in \GX~can be a factor of $\sim$2 smaller than that in XTE~J1550--564 (see Table~\ref{tbl:intrvl}; also \citealt{Orocz2011}).
Geometrically thicker outer parts of the disk may also increase the amount of irradiation flux received.
Then, the light bending effects in the vicinity of a Kerr black hole substantially increase the intensity of the illuminating flux \citep{Suleimanov2008}. 
There is evidence that the black hole spin of \GX~can be as large as $a \approx 0.935$ (\citealt{Reis2008}, \citealt{Miller2008}; but see also \citealt{Kolehmainen2010,Kolehmainen2011} who give an upper limit of 0.9), compared to $a \approx 0.5$ in XTE~J1550--564 \citep{Steiner2011}.
It is possible that a combination of these factors leads to a higher SS temperatures inferred in \GX.
Furthermore, the extinction in XTE~J1550--564 assumed by \citet{Poutanen2014} might have been underestimated; a higher $A_V$ would result in a larger temperature bringing it closer to the values measured in \GX.

The Q states of \GX~appear to deviate from those observed in XTE~J1550--564.
While for XTE~J1550--564 these points lie on the blackbody track, for \GX~they lie below the blackbody curve.
It is known that the contribution of the companion star is small even in quiescence (up to 50 per cent to the $J$ and $H$ filters, \citealt{Heida2017}), and the red spectrum of the K-type companion cannot explain the shift towards the blue part of the CMD, as well as substantial variability of colors in \GX. 
Other LMXBs are known to exhibit quiescent flickering, but the magnitude of quiescent variability of \GX~and significant changes in observed colors exceed that measured for other sources \citep{Zurita2003}.
A possible explanation for this behavior is the emission from the spiral arms, hotspots and hot line, flares triggered by the magnetic field reconnection in the vicinity of the disk, and the presence of edges and lines in the quiescent spectra \citep[see e.g.][]{Zurita2003,Cherepashchuk19,Baptista2020}.

We find that the regular and failed outbursts occupy the same regions in the CMD.
Though the regular outbursts tend to be more luminous during the RHS, the position of points of 2004--2005 outburst intersects with brightest parts of the analysed failed outbursts.
We thus conclude that the peak ONIR brightness has no predictive power for the failed outbursts.
The reason for the similarity of ONIR brightness of regular and failed outbursts is not clear from the light curve and CMD analysis.
If the failed outbursts differ from the regular ones by the mass transfer rate, then the similar brightness during the RHS can be explained by a weak dependence of the non-thermal component on the accretion rate.
Because it dominates in the ONIR, the difference between regular and faint outbursts is minor.

\subsection{ONIR vs X-ray correlation}
\label{sec:ONNIRvsXRAY}

\begin{figure*}
    \centering 
        \includegraphics[keepaspectratio, width = 0.8\linewidth]{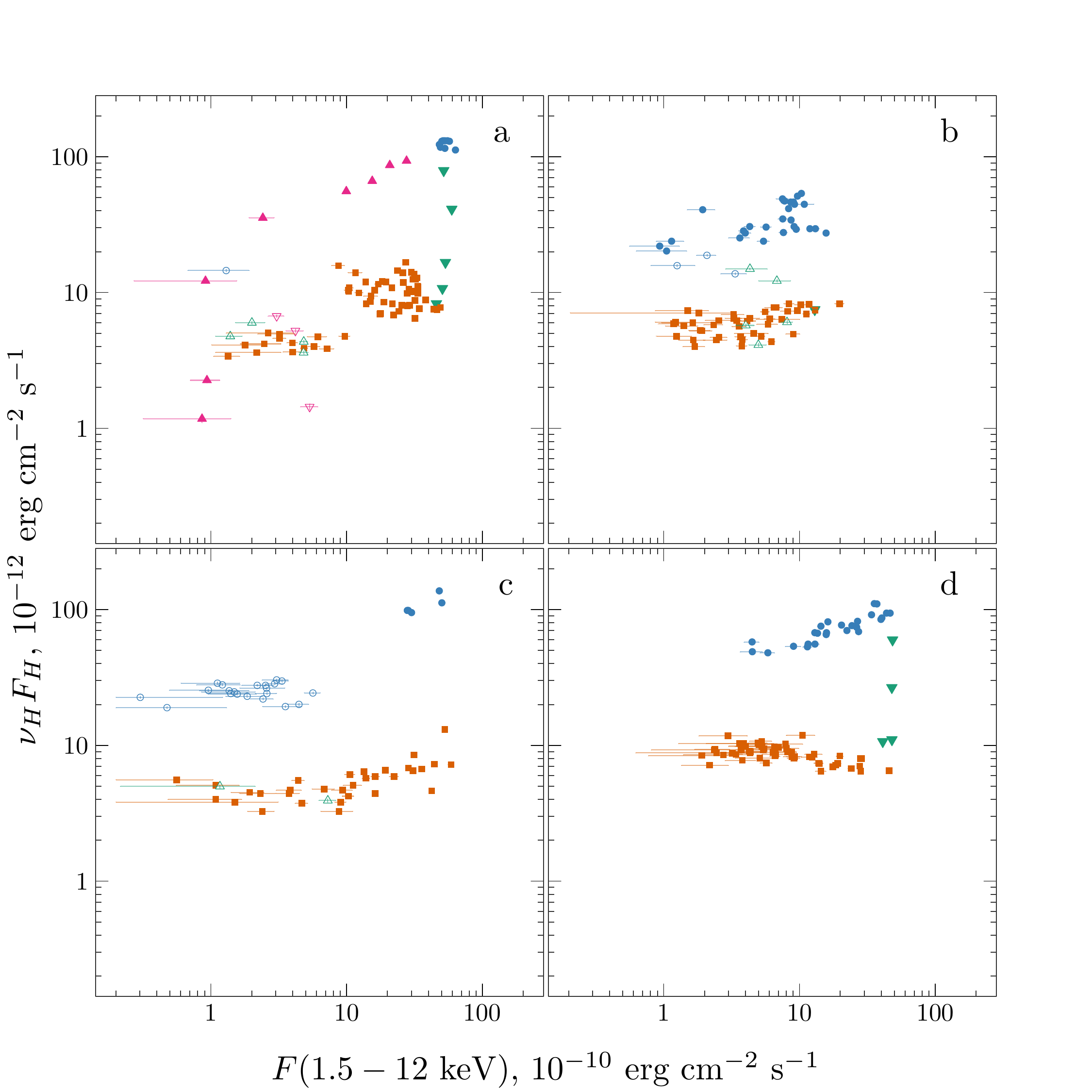}
        \caption{Quasi-simultaneous ONIR $\nu_H F_H$ versus X-ray $1.5 - 12~\mrm{keV}$ flux diagram. The X-ray data are taken from all three ASM bands. The colors and symbols are the same as in Fig.~\ref{fig:CMD_LC_01} and errors are 1$\sigma$. (a) 2002--2003, (b) 2004--2005, (c) 2007, and (d) 2010--2011 outbursts.}
        \label{fig:FvsFx}
\end{figure*}

\begin{figure}
    \centering 
        \includegraphics[keepaspectratio, width = 0.9\linewidth]{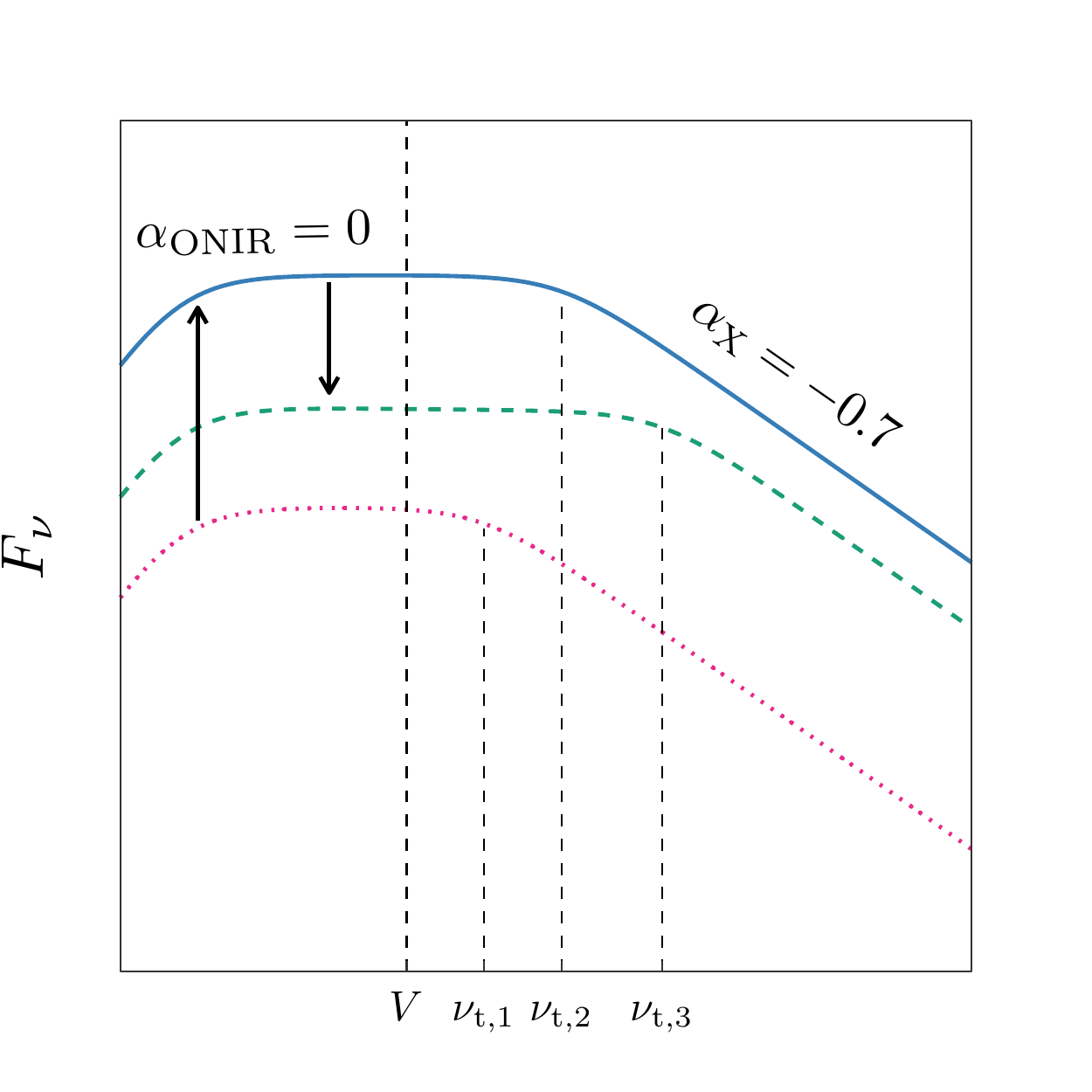}
        \caption{Schematic representation of the evolution of \GX~spectra during the rising phase hard state and HtS transition of an outburst. 
        The dotted pink line is the RPh hard spectrum, the solid blue line is the RHS, and the dashed green line is HtS transition. 
        Vertical black arrows indicate the transition sequence. 
        Characteristic turnover frequency of each spectrum and the $V$ filter frequency are shown with the vertical dashed lines. 
        Characteristic spectral slopes in the ONIR and X-ray ranges are shown with  $\alpha_\mrm{ONIR}$ and $\alpha_\mrm{X}$. 
        }
        \label{fig:sketches}
\end{figure}
    
Black hole LMXBs show strong non-linear correlation between radio, ONIR and X-ray fluxes. 
The radio--X-ray correlation was first documented for Cyg X-1, \GX~and V404 Cyg \citep{Brocksopp99,Corbel00,Gallo2003} and later was confirmed on a much larger sample of Galactic LMXBs \citep[including \GX~during multiple outbursts, see][]{Corbel2013, Islam2018}. 
The global ONIR--X-ray correlation, which naturally arises from the different emission mechanisms \citep[][]{vanParadijs1994, Gallo2003}, was observed for a number of black hole LMXBs \citep{Russell2006} and for the 2002--2003, 2004--2005 and 2007 outbursts of \GX, in particular \citep[see][]{HBMB05, Coriat2009}.
Here we expand the sample to include the 2010--2011 outburst.

We use the {\it RXTE}/ASM flux as a proxy for the X-ray luminosity. 
The observed flux, however, cannot be simply transformed into the bolometric luminosity, as the latter depends on the spectral shape.
The bolometric correction is large in the hard state and the quiescence  as most of the emission is radiated at energies above the ASM passband.
We use the observed $H$-band flux as a proxy for the luminosity of the non-thermal component.
Fig.~\ref{fig:FvsFx} shows the relation between the $H$-band and X-ray fluxes 
for the regular outbursts \citep[see also][]{Zdziarski2002}. 
The source moves in the clockwise direction on this diagram with the hysteresis pattern being similar to that found by \citet{Coriat2009}. 

In the beginning of the SS, the $H$-band flux is nearly constant, with a hint of anti-correlation with the X-ray flux in the last outburst (Fig.~\ref{fig:FvsFx}d), which is likely caused by the changing bolometric correction.
Such anti-correlation between ONIR and X-ray fluxes is atypical for \GX~and other LMXBs, for which HS and SS correlations have very similar slopes \citep[Fig.~\ref{fig:FvsFx}a,b,c; also][]{Russell2006, Russell2007, Coriat2009}. 
The RHS of the 2010--2011 outburst (Fig.~\ref{fig:FvsFx}d) seems to have a slope, which is not as steep as observed in other outbursts, with 
$\partial\log(\nu_H F_H) / \partial\log F(\mathrm{ASM})\approx 0.3$, as compared to $\approx 0.4$ for the 2002--2003 and 2007 events.

The relation between ONIR and X-ray fluxes can be used to put constraints on the models of spectral formation in accreting black holes. 
The scaling of the ONIR and X-ray fluxes with the mass accretion rate was used to predict the observed ONIR -- X-ray correlation in the jet-dominated scenario \citep{Russell2006}.
Here we suggest an interpretation of the HS correlation  in the framework of the hot accretion flow model \citep{Veledina2013, Poutanen2014a}.  
The schematic representation of the spectrum in the ONIR--X-ray range is shown in Fig.~\ref{fig:sketches}. 
For a broken power-law spectrum, we can relate the  ONIR ($L_\mrm{o}$) and X-ray ($L_\mrm{x}$) luminosities as: 
\begin{equation}
    \frac{L_\mrm{o}}{L_{\nu_\mrm{t}}} = \left(\frac{\nu_\mrm{o}}{\nu_\mrm{t}}\right)^{\alpha_\mrm{ONIR}}, \qquad 
    \frac{L_\mrm{x}}{L_{\nu_\mrm{t}}} = \left(\frac{\nu_\mrm{x}}{\nu_\mrm{t}}\right)^{\alpha_\mrm{X}},
    \label{eq:SPEC_SHAPE}
\end{equation}
where $\nu_\mrm{t}$ is the (break) turnover frequency, $\alpha_\mrm{ONIR}$ and $\alpha_\mrm{X}$ are spectral slopes and $\nu_\mrm{o}$ and $\nu_\mrm{x}$ are characteristic frequencies in the ONIR and X-ray ranges, respectively. 
The ratio of the X-ray to ONIR luminosity is then
\begin{equation}    \label{eq:Lx_Lv}
    \frac{L_\mrm{x}}{L_\mrm{o}} = \left(\frac{\nu_\mrm{x}}{\nu_\mrm{t}}\right)^{\alpha_\mrm{X}}\left(\frac{\nu_\mrm{o}}{\nu_\mrm{t}}\right)^{-\alpha_\mrm{ONIR}} =
    \frac{\nu_\mrm{x}^{\alpha_\mrm{X}}}{\nu_\mrm{o}^{\alpha_\mrm{ONIR}}}\nu_\mrm{t}^{\alpha_\mrm{ONIR} - \alpha_\mrm{X}} .
\end{equation}
Therefore the slope of the optical--X-ray correlation is 
\begin{equation}     \label{eq:gamma_beta}
    \gamma \equiv \frac{\partial \log L_\mrm{o} }{\partial \log L_\mrm{x}} = 1 - \beta\left(\alpha_\mrm{ONIR} -\alpha_\mrm{X}\right), 
\end{equation}
where $\beta  \equiv \partial \log\nu_\mrm{t} / \partial \log L_\mrm{x}$.

We take $\alpha_\mrm{X} \approx -0.7$ and $\alpha_\mrm{ONIR} \approx 0$ (average of four regular outbursts, see Sect.~\ref{sec:results} and  Fig.~\ref{fig:SPEC_1_4}). 
The turnover frequency scales with the magnetic field $B$ and the Thomson optical depth $\tau$ across the flow as
\begin{equation} 
\nu_\mrm{t} \propto B^{\frac{p+2}{p+4}} \tau^{\frac{2}{p+4}} ,
\end{equation}
where $p$ is the power-law slope of the electron distribution.
Assuming density scaling with the accretion rate $\rho \propto \dot{m}$ (i.e. $\tau\propto \dot{m}$ too) and equipartition magnetic field $B^2 \propto \rho$ (assuming constant temperature), we arrive at 
\begin{equation} 
\nu_\mrm{t} \propto \dot{m}^\frac{p+6}{2(p+4)} .
\end{equation}

The bolometric luminosity scales with the accretion rate as $\dot{m}$ or as $\dot{m}^2$ for the cases of hot, radiatively efficient \citep[e.g.][]{Bisnovatyi-Kogan97} or inefficient (advection-dominated) accretion flow \citep[e.g.][]{NarayanYi95}, respectively.  
For these two cases we get 
\begin{equation} 
\beta_{\rm bol} \equiv \frac{\partial \log\nu_\mrm{t}}{\partial \log L_\mrm{bol}} = \frac{p+6}{2(p+4)} \quad \mbox{or} \quad \beta_{\rm bol} = \frac{p+6}{4(p+4)}.
\end{equation}
Taking $p=4$ to satisfy the slope of the MeV emission observed from Cyg X-1 \citep{McConnell02}, we get $\beta_{\rm bol} = 0.63$ and 0.31 for  the two cases. 
Assuming $\beta=\beta_{\rm bol}$ we get $\gamma \approx 0.56$ and 0.78, respectively. 
This range of $\gamma$ is consistent with the observed slope of the optical--X-ray correlation $\gamma_\mrm{obs} \approx 0.6$ obtained in \citet{Russell2006} for a number of black hole LMXB sources.
We observe $\gamma_\mrm{obs} \approx 0.4$ (similar to $\gamma_\mrm{obs} \approx 0.48$ in \citealt{Coriat2009}).
Such slopes require $\beta \approx 0.8-0.9$, which is hard to achieve.
However, the underestimated value of $\gamma$ can be caused by the variable bolometric correction. 
If it is positively correlated with the luminosity  (see \citealt{Zdziarski2004,Koljonen2019})\footnote{We note that bolometric correction can increase towards lower luminosities, because the rising electron temperature \citep{Veledina2013} results in a larger fraction of flux to emerge above 200 keV as well as in the ONIR band.} then  
\begin{equation} 
\beta = \beta_{\rm bol}\ \frac{\partial \log L_\mrm{bol} }{\partial \log L_\mrm{x}} > \beta_{\rm bol} , 
\end{equation}
bringing $\gamma$ closer to the observed range.

\subsection{Origin of the non-thermal ONIR component}

The nature of the red non-thermal component seen in LMXBs during the ONIR flares is debated.
Two main candidates are proposed: synchrotron emission from the jet or from the hybrid hot accretion flow \citep[see reviews by][]{Uttley2014, Poutanen2014a}.
In some cases the ONIR soft spectrum is consistent with the optically thin synchrotron emission, and has been modeled with the jet \citep[e.g. MAXI~J1836--194; ][]{Russell2014,Peault2019}. 
In other cases, when the disk-subtracted non-thermal component had a hard spectrum \citep{Poutanen2014}, or when the extrapolation of the radio continuum significantly underestimates ONIR fluxes \citep[e.g. SWIFT~J1753.5--0127; ][]{Chiang2010, Kajava2016}, the excess ONIR emission has been attributed to the hot accretion flow.
A way to discriminate between the two models is to track the behavior of non-thermal component during the ONIR state transition.
In a simple jet scenario, the spectrum from radio to optical wavelengths can be described by a broken power law \citep{Blandford1979}, with partially-absorbed and optically thin parts.
During the HtS transition the synchrotron break frequency is expected to decrease, as it scales inversely-proportional to the inner disk radius \citep{Heinz2003}. 
In the hot flow model, the spectrum from infrared to X-rays is expected to have two breaks: the lower-frequency break between the partially-absorbed and fully self-absorbed parts (it corresponds to the extent of the flow), and the higher-frequency break between the partially-absorbed and synchrotron Comptonization parts \citep{Veledina2013}.
Under the simplest assumptions, the state transition is accompanied by the collapse of the flow outer parts, and so the lower-frequency break is expected to increase, while the spectral slope of the partially self-absorbed part can remain the same. 
The reverse trends are expected in the same scenarios for the StH transitions.

Unlike expected in these two simplest scenarios, we find that the non-thermal component has a constant spectral shape during the HtS and StH transitions, and only its normalization decreases (see lower panels in Fig.~\ref{fig:SPEC_1_4}).
We propose that the aforementioned evolution can be understood in terms of the hot-flow model, if both the energy release and the electron number density increase when the accretion rate increases during RPh.
This then leads to the increase of total luminosity and the increase of the higher break frequency.
As the accretion rate increases, the synchrotron emission, produced at the same physical radius in the flow, would shift to higher frequencies (as the frequency of self-absorption increases) and become brighter (see changes between dotted and solid lines and between turnover frequencies $\nu_\mrm{t,1}$ and $\nu_\mrm{t,2}$ in Fig.~\ref{fig:sketches}).
In this model, the RHS is described by the increase of ONIR luminosity with a nearly constant spectral shape (as the latter is determined by the distribution of parameters over the radius, rather than by their absolute values).
At the HtS transition, when the energy release is shifting from the hot accretion flow to the geometrically thin accretion disk, the  energy deposited in the hot flow decreases, while the particle density may still increase. 
This results in a higher break frequency, but smaller overall ONIR luminosity (see changes between solid and dashed lines and between turnover frequencies $\nu_\mrm{t,2}$ and $\nu_\mrm{t,3}$ in Fig.~\ref{fig:sketches}).

The hot-flow model is capable of explaining both (nearly) constant colors of the non-thermal component during the RHS, as well as the constant spectral shape at decreasing ONIR luminosity during the HtS state transition if the hot flow extent during the observed transitions has not decreased. 
This would guarantee that the emission in the $H$ filter appears in the partially self-absorbed part. 
Thus in contrast to the expectations of the original scenario the data can be described by the hot-flow model if there is no collapse of the outer parts of the flow. 
We suggest that the survival of the synchrotron-emitting plasma with the inwards-moving inner edge of the disk can be seen as the transition from the hot flow (within the boundaries of the cold disk) to the corona atop of the accretion disk.
This scenario is supported by the StH transition and the DHS, where we observe the same evolution happening in reverse. During the StH the spectra are nearly flat and their shapes do not change (see Figs.~\ref{fig:SPEC_1_4}a$^*$,c$^*$,d$^*$,f$^*$), indicating that the $H$ filter remains in the self-absorbed part and we observe a reverse process of transition from the hot corona atop of the disk to the hot flow, as the inner edge of the disk moves outward.

The shape of the ONIR spectra in the simplest jet scenario depends on the break frequency, which may lie beyond the $V$-band ($>5\times 10^{14}$ Hz, see \citealt{Coriat2009,Dincer2012,Buxton2012}), or in the mid-IR ($10^{13-14}$ Hz, see e.g. \citealt{Corbel2002,Gandhi2011}). 
The problem is that estimates of the break frequency were obtained for different outbursts using different estimates for the extinction (see \citealt{HBMB05}, \citealt{Maitra2009}, and \citealt{Buxton2012}). 
In our interpretation of
the quasi-simultaneous mid-IR data obtained with \textit{WISE} \citep{Gandhi2011}, there is likely another highly variable soft non-thermal component, which contributes to the mid-IR in addition to the power law.
\GX~shows two distinct mid-IR spectral profiles: a low-flux flat one, which agrees with the power-law-like disk-subtracted ONIR spectrum, and a high-flux soft one, which drops sharply between $W1$ and $H$ filters (see Fig.~\ref{fig:broadband}).
If we interpret this additional soft component as coming from the jet, the jet break frequency is  about $10^{14}$~Hz (see Table~\ref{tbl:mdl_fit_param}). 
This implies that the jet can dominate mid-IR fluxes, but as soon as the jet spectral slope (as determined from the radio data) changes to negative, the hot flow becomes the dominant source of the mid-IR emission, providing a baseline flux level in the mid-IR and ONIR filters.
The shape and evolution of the ONIR spectra are then determined by the sum of the thermal disk and flat hot-flow components with little to no contribution from the jet.

The fast variability observed in \GX~can shed more light on the nature of the ONIR emission. Within the hot-flow model, two components contribute to the variability in ONIR: synchrotron emission from the hot flow and reprocessing of the X-rays in the accretion disk. 
In this model  \citep{VPV11}, a precognition dip in the optical-X-ray CCF is explained by anti-correlation of the ONIR synchrotron and the X-ray Comptonization components. 
The reprocessing produces a positive peak at positive lags. 
Together they can produce a typical CCF observed in many BH X-ray binaries \citep[see][]{Poutanen2014a} including \GX~\citep{GMD08, Gandhi2010}. 
The IR/X-ray correlation, however, appears to depend on the type and phase of the outburst. The CCF obtained for the 2008 failed outburst (low-flux hard state) shows no precognition dip and only slight asymmetry \citep{CMO10}, while CCF of the RHS of the 2010--2011 outburst features a precognition dip \citep{Kalamkar2016}. Although it is possible to explain this behavior within jet-only paradigm \citep[see e.g.][]{Malzac2018}, it is likely that both jet and accretion flow produce different features of the IR/X-ray CCF (see also Section~\ref{sec:IR}).

\section{Conclusions}
\label{sec:conclusions}
 
We analyzed the 2002--2011 ONIR light curves of \GX. 
We used the ONIR data to separate the states of the source and compared the estimated transition dates with those obtained from the X-ray spectra. 
We measured the typical durations of various outburst stages: the HtS state transitions take typically a week, while the StH state transitions last twice longer. 
The duration of the hard state at the decaying stage is about four weeks. 

Using the SS data from four regular outbursts, we determined the interstellar extinction towards the source, which allowed us to measure the intrinsic ONIR spectral slopes during different phases of the outbursts.
We showed that during the outbursts the source passes through the same areas on the CMD, giving us an opportunity to construct a template where different outburst phases are identified.

During the SS, the object is found to follow the blackbody model with a constant normalization. 
During the HS, a red non-thermal component dominates the $H$ band, with the flux being higher at the rising phase than at the decaying phase. 
We find that the end of the HtS transitions corresponds to the blackbody temperature is 30–40 kK, while at the start of the reverse transition the temperature is $\approx$20~kK. 
In quiescence, the spectrum of \GX~is bluer than the blackbody model with the same normalization, which can be explained either by the presence of strong spectral features (edges), or by the decrease of the apparent emitting area.
The corresponding disk temperatures are well above the hydrogen ionization limit.
The variability of the source in this state exceeds the measurements errors and its nature is uncertain.

We traced the evolutionary tracks of the failed outbursts on the CMD and found that their HS show the same colors as the HS of regular outbursts, but lower ONIR fluxes. 
The decay of a failed outburst follows the same track as the decaying phase of a regular outburst. 
These properties make it difficult to predict the failed outburst at early stages using ONIR data alone. 
We found that a bright RHS, $V < 16$\,mag, is a sign of a regular outburst.

Finally, we investigated the behavior of the non-thermal component, which dominates the ONIR fluxes during the HS and the state transitions.
For this, we subtracted the contribution of the thermal component (disk) using the extrapolated SS light curves.
We found that the luminosity of the non-thermal component evolves with a nearly constant power-law spectrum with energy spectral index $\alpha \approx 0$ (where $F_{\nu}\propto\nu^{\alpha}$).
We proposed that this evolution can be explained by a model where the non-thermal component originates from the synchrotron emission of the hot accretion flow. 
However, for this model to work, the extent of the hot medium should not decrease at the state transitions, but instead the hot flow should transforms into the hot corona atop of the accretion disk as the inner disk radius moves inward during the HtS transition. 
In this case, the $H$-filter remains in the partially self-absorbed region implying constant spectral shape. 
The reverse process may occur during the StH transition. 
Using radio ATCA and mid-IR \textit{WISE} data contemporaneous with the available SMARTS data, we also showed 
that an additional strongly variable component, likely associated with the jet, is required in the mid-IR band. 
The spectrum of this component has a sharp cutoff below $10^{14}$\,Hz.  
We also detected a low-frequency cutoff in the hot flow component at $\sim 10^{13}$\,Hz, which can be used to estimate its size of about $3\times 10^9$\,cm.
We conclude that at least three components (the disk, the  hot flow and  the jet) are required to explain the broad-band mid-IR to ONIR spectra of \GX\ at different stages of its outburst.

\section*{Acknowledgements}

We thank the anonymous referees for valuable comments and suggestions, which helped improve the manuscript.
AV acknowledges support from the Academy of Finland grant 309308.
VFS was supported by the DFG grant WE 1312/51-1, the German Academic Exchange Service (DAAD) travel grants 57405000 and 57525212 and by the Magnus Ehrnrooth Foundation travel grant. 
This research was also supported by the Academy of Finland travel grants 317552 and 331951 (IAK, JP).
This paper has made use of SMARTS optical/near-infrared light curves. 
AV thanks the International Space Science Institute (ISSI) in Bern, Switzerland for support. 

\bibliographystyle{aa}
\bibliography{references}

\appendix
\section{Separation of the states}
\label{Appendix}

We employ a synthetic algorithm in order to distinguish between different states of the source. We distinguish 8 different phases of an outburst (including rapid transitions between states). Each phase can be identified quite easily on either color-magnitude or light curve plots. However, it can be difficult to determine exact dates at which the transitions occur. The simultaneous analysis of four different outbursts requires an objective method to determine transition dates in all of the outbursts without human intervention.

We developed an algorithm to fit a model to the transition data. We subdivide transitions into two categories: HtS and StH transitions, which share similar S- or Z-shaped  profile, and the transitions to/from quiescence. The evolution of the ONIR fluxes of \GX~in the vicinity of the HtS and StH can usually be  well approximated by an exponential rise or decay, which corresponds to a linear trend in magnitudes. A simple a sigmoid function fits two different linear trends before and after the transition and matches the intermediate S-shaped evolution:
\begin{equation}
    \label{eq:sigmoid}
    S(t) = 
        \frac{m_1 + \mu_1\left(t - t_0\right)}{1 + \exp\left(\frac{t-t_0}{T}\right)} + 
        \frac{m_2 + \mu_2\left(t - t_0\right)}{1 + \exp\left(-\frac{t-t_0}{T}\right)} , 
\end{equation}
where $m_1$, $\mu_1$, $m_2$, $\mu_2$ are parameters of the linear functions, $t_0$ is a turnover point, at which $S(t_0) = 1/2(m_1 + m_2)$, and $T$ is the characteristic timescale of the transition. We then fit this model to the HtS and StH transitions using the data from the $H$ filter light curve. The reason we chose the $H$ filter is that the non-thermal component contributes more to this filter, making it easier to determine the turnover points.

\begin{figure}
\centering
       \includegraphics[keepaspectratio, width = 0.95\linewidth]{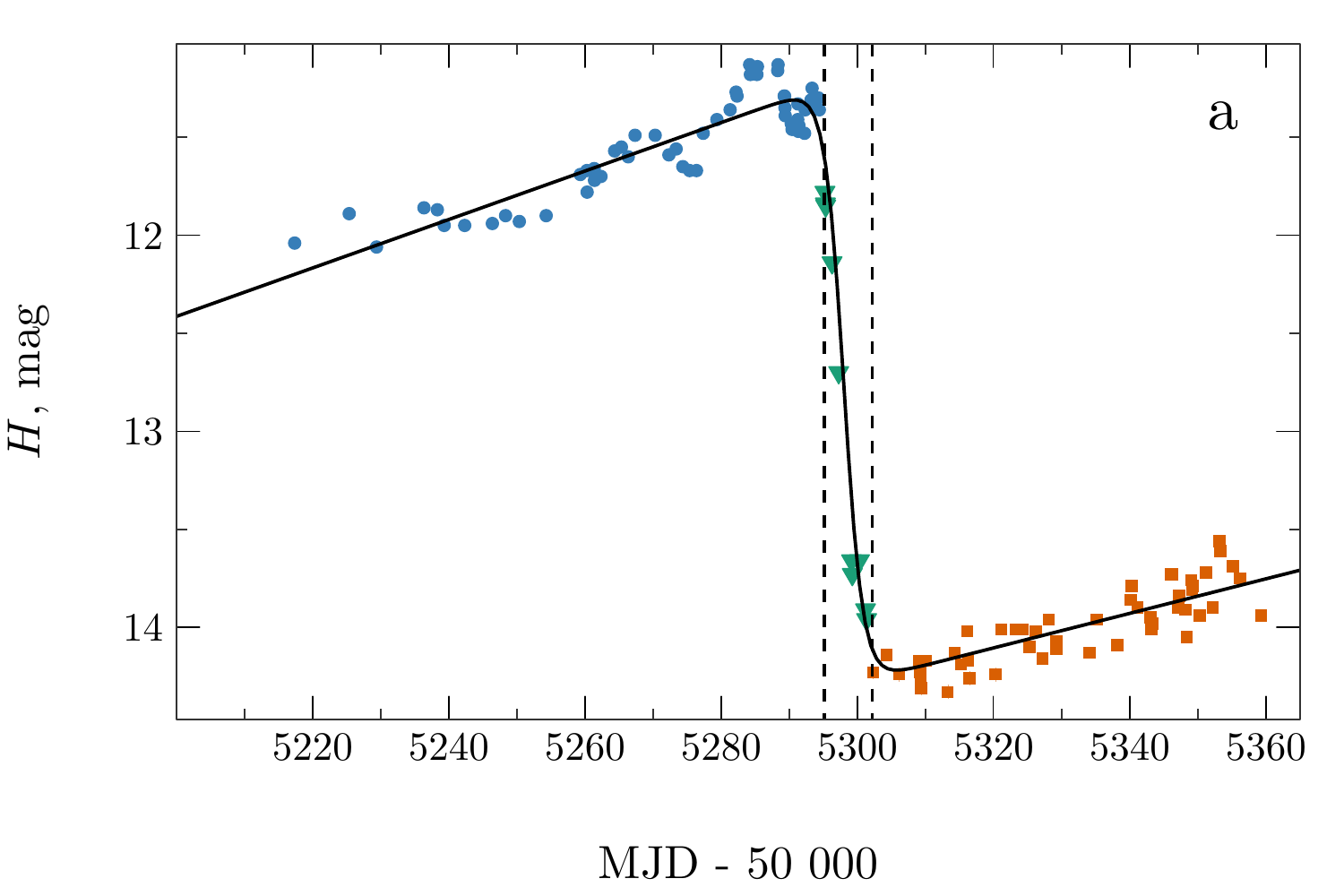}
       \includegraphics[keepaspectratio, width = 0.95\linewidth]{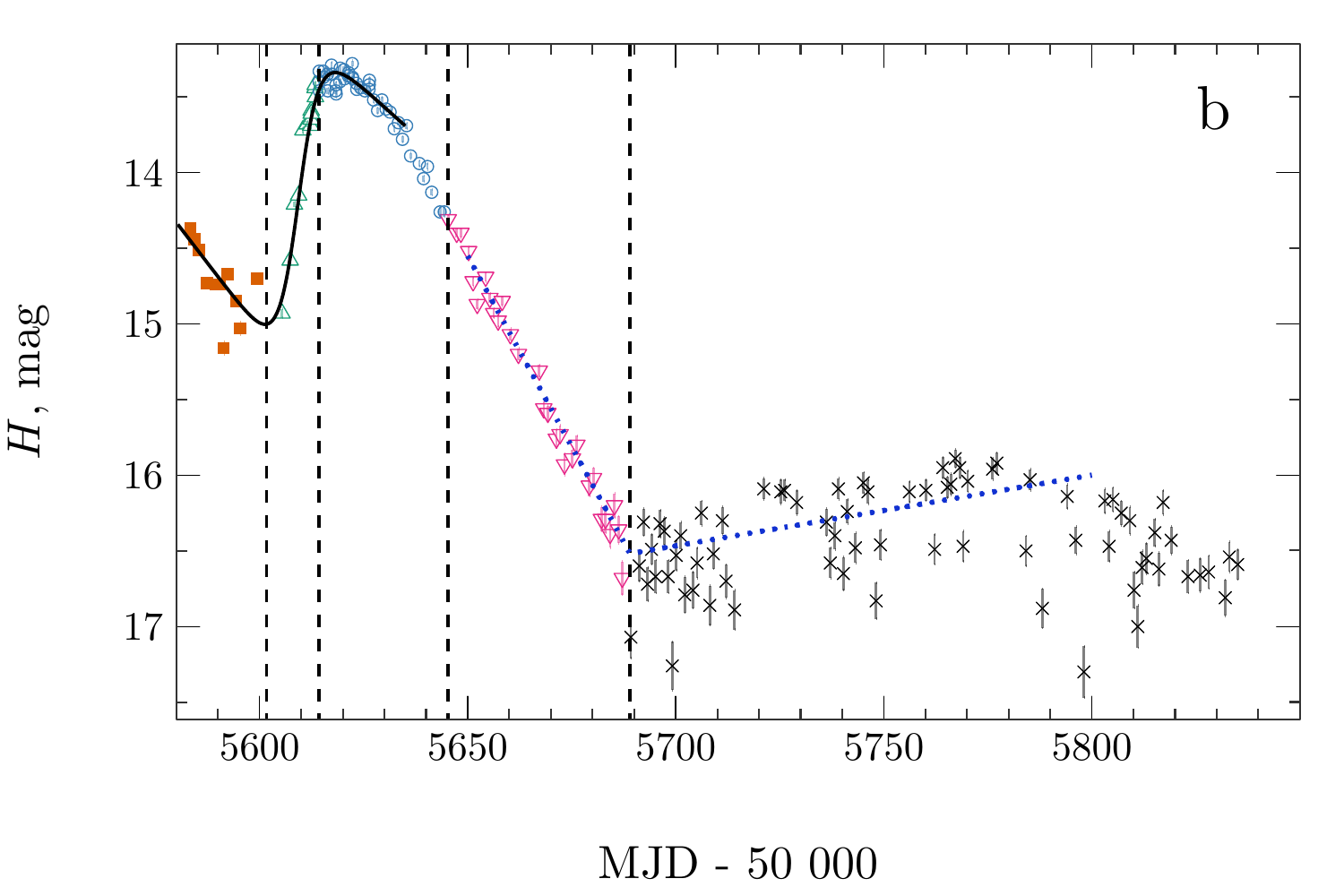}
    \caption{Example of the phase separation during (a) rise and (b) decay of the 2010--2011 outburst. 
    Colors and symbols are same as in Fig.~\ref{fig:TEMPL}. 
    The solid black line and dotted blue line (panel b)  show the fitted models given by Eq.~(\ref{eq:sigmoid}).
    Vertical dashed lines correspond to the boundaries between different phases.} 
    \label{fig:SEP_1}
\end{figure}

As a result, the transition boundaries can be universally expressed in terms of $t_0$ and $T$. We define HtS start date as $t_0 - 2T$ and end date as $t_0 + 3T$. Similarly, for the StH transition the start date is $t_0 - 3T$ and the end date is $t_0 + 2T$. The choice of the offset factors is influenced by the slopes of the trends before and after the transition. For example, at $t = t_0 - 3T$ the contributions of the first and second trends are 5 and 95\%, respectively, while at $t = t_0 - 2T$ the trends contribute 12 and 88\%, respectively. As a result of such choice, the duration of the StH or HtS transitions is $5T$ and can be directly obtained from the fitted model parameters.  An example of such outburst separation is shown in Fig.~\ref{fig:SEP_1}(a).

The boundaries between the decaying phase and the quiescence were determined in a similar manner. Due to the differences in the quiescent state baseline fluxes between outbursts, there is no common lower limit applicable to all of the outbursts simultaneously, which can be used to determine the end of the decay. Therefore, we fit similar model to the decaying phase and quiescence data, for which we fix the parameter $T$ at $\frac{1}{3}$\,d, corresponding to the very rapid change of the fitted model from one trend to another. We then use the turnover point $t_0$ as a formal definition of the boundary between transition and quiescence (see Fig.~\ref{fig:SEP_1}b, dotted blue line). 

Our algorithm was also applied to the failed outbursts. We fitted our model to the 2006 and 2009 events and adopted $t_0 - 2T$ and $t_0 + 2T$ as the transition boundaries. For the 2008 failed outburst, we used the turnover points $t_0$ to subdivide the light curve into 4 parts. During this event \GX~showed extremely erratic behavior with no clear manifestations of the non-thermal component, therefore the phase separation is rather arbitrary and is only required for comparison to other failed outbursts.

In order to compare the decaying phase of the failed and regular outbursts, we split the interval between the end of the SS and the beginning of the quiescence in half and mark the second half as DPh (see Fig.~\ref{fig:SEP_1}b, pink open triangles). We also apply the algorithm to the rising phase of the 2002--2003 outburst, obtaining the starting date of the RPh. 

Our algorithm is similar to the one used by \citet{BB04} and is relatively insensitive to the data selection effects, as we eliminate the process of by-eye separation of data into groups prior to fitting \citep{KDT13}. As a result, this method can be used to study and compare transition dates of the outbursts observed in other LMXB transients.

\end{document}